\documentclass{bmcart}

\usepackage[utf8]{inputenc} %unicode support
\usepackage{amsfonts,amsmath,amssymb,bm,eucal,slashed}
\usepackage{lscape}
\usepackage{rotating}
\usepackage{url}
\usepackage[english]{babel}
\usepackage[T1]{fontenc}
\usepackage{ae,aecompl}
\usepackage{graphicx,graphics,epsfig,epstopdf,color,psfrag,subfigure}
\usepackage{hyperref}

\startlocaldefs
\endlocaldefs

\begin{document}

\begin{frontmatter}

\begin{fmbox}
\dochead{Regular Article}

\title{Complex networks and public funding: the case of the 2007-2013 Italian program}

%%%%%%%%%%%%%%%%%%%%%%%%%%%%%%%%%%%%%%%%%%%%%%
%%
%% Enter the authors here
%%
%% Specify information, if available,
%% in the form:
%%   <key>={<id1>,<id2>}
%%   <key>=
%% Comment or delete the keys which are
%% not used. Repeat \author command as much
%% as required.
%%
%%%%%%%%%%%%%%%%%%%%%%%%%%%%%%%%%%%%%%%%%%%%%%

\author[
   addressref={aff1},				% id's of addresses, e.g. {aff1,aff2}
   corref={aff1},					% id of corresponding address, if any
   noteref={n1},				% id's of article notes, if any
   email={stefano.nicotri@ba.infn.it}	% email address
]{\inits{S}\fnm{Stefano} \snm{Nicotri}}
\author[
   addressref={aff2,aff3},
   email={eufemia.tinelli@uniba.it}
]{\inits{E}\fnm{Eufemia} \snm{Tinelli}}
\author[
   addressref={aff1,aff2},
   email={nicola.amoroso@ba.infn.it}
]{\inits{N}\fnm{Nicola} \snm{Amoroso}}
\author[
   addressref={aff4},
   email={elena.garuccio@gmail.com}
]{\inits{E}\fnm{Elena} \snm{Garuccio}}
\author[
   addressref={aff1,aff2},
   email={roberto.bellotti@uniba.it}
]{\inits{R}\fnm{Roberto} \snm{Bellotti}}

%%%%%%%%%%%%%%%%%%%%%%%%%%%%%%%%%%%%%%%%%%%%%%
%%
%% Enter the authors' addresses here
%%
%% Repeat \address commands as much as
%% required.
%%
%%%%%%%%%%%%%%%%%%%%%%%%%%%%%%%%%%%%%%%%%%%%%%

\address[id=aff1]{					% unique id
  \orgname{Istituto Nazionale di Fisica Nucleare - Sezione di Bari}, 
  \street{via Orabona 4},
  \postcode{I-70125}
  \city{Bari},
  \cny{Italy}
}

\address[id=aff2]{
  \orgname{Dipartimento Interateneo di Fisica \emph{``M.~Merlin''}, Universit\`a degli Studi di Bari \emph{``A.~Moro''}},
  \street{via Orabona 4},
  \postcode{I-70125}
  \city{Bari},
  \cny{Italy}
}

\address[id=aff3]{
  \orgname{Comune di Bari - Ripartizione Innovazione Tecnologica, Sistemi Informativi e TLC},
  \street{Corso Vittorio EmanueIe II, 143},
  \postcode{I-70122}
  \city{Bari},
  \cny{Italy}
}

\address[id=aff4]{
  \orgname{Dipartimento di Scienze Fisiche dell'Ambiente e della Terra, Universit\`a degli Studi di Siena 1240},
  \street{via Roma 56},
  \postcode{I-53100}
  \city{Siena},
  \cny{Italy}
}

%%%%%%%%%%%%%%%%%%%%%%%%%%%%%%%%%%%%%%%%%%%%%%
%%
%% Enter short notes here
%%
%% Short notes will be after addresses
%% on first page.
%%
%%%%%%%%%%%%%%%%%%%%%%%%%%%%%%%%%%%%%%%%%%%%%%

\begin{artnotes}
%\note{Sample of title note}  % note to the article
\note[id=n1]{Corresponding author}	% note, connected to author
\end{artnotes}

\end{fmbox}  % comment this for two column layout

\begin{abstractbox}

\begin{abstract}
In this paper we apply techniques of complex network analysis to data sources representing public funding programs and discuss the importance of the considered indicators for program evaluation. 
Starting from the Open Data repository of the 2007-2013 Italian Program \emph{Programma Operativo Nazionale ``Ricerca e Competitivit\`a''} (PON R\&C), we build a set of data models and perform network analysis over them. 
We discuss the obtained experimental results outlining interesting new perspectives that emerge from the application of the proposed methods to the socio-economical evaluation of funded programs.
\end{abstract}

\begin{keyword}
\kwd{public funding}
\kwd{open data}
\kwd{complex networks}
\kwd{program evaluation}
\end{keyword}

% MSC classifications codes, if any
%\begin{keyword}[class=AMS]
%\kwd[Primary ]{}
%\kwd{}
%\kwd[; secondary ]{}
%\end{keyword}

\end{abstractbox}

%\end{fmbox}						% uncomment this for twcolumn layout

\end{frontmatter}

\section{Introduction}\label{sec:introduction}

Since the last years of the past century, the importance of basing policies on evidence, data, and analysis has quickly spread all over the world.
The Evidence-Based Policy movement \cite{Smith1996,nutley2000works,solesbury2001evidence,SPS:110009,AUPA:AUPA728} has grown enormously, and mainly all public administrations are now focused on maximising utility and show a pragmatic problem-solving approach to socio-economical issues \cite{doi:10.1080/01442872.2012.695945}.
In this respect, the evaluation of public funding programs is a field of great interest for policymakers and economists.
Politicians and technicians need to estimate the impact that funding has on life and society, in order to address future programs and to modify their decisions.
Many standard and advanced statistical methods are commonly used for this purpose, such as linear/nonlinear regressions, Bayesian inference, machine learning, data mining, and so on.
In this paper we suggest new indicators, coming from network analysis, that can help underlining in a quantitative way important effects that are not usually considered, being them outside the domain of investigation of standard statistical tools.
This does certainly not mean that program evaluation cannot be performed without including network analysis, but that valuable insight about public funding programs could hopefully be inferred from such techniques, in order to help increasing objectivity of the extracted results.
Recently, a growing interest towards complex network analysis applied to evaluation can be seen both in literature \cite{PUAR:PUAR045,NetworksandInterOrganizationalManagement,Penuel01122006,Horelli01042009,Ploszaj1,EPJB12292} and institutional reports \cite{agendadigitalepuglia2020}.
The indicators we suggest can be used by experts in program evaluation for their analyses, giving them the opportunity of considering and quantitatively measuring important features of the funding programs, such as relations between the actors involved in them. 
Social network analysis is a particularly suitable tool to extract information about relations among the different components of a system.
Investigating the relations between the actors participating to a program could be of interest, since can \emph{e.g.} show structural contradictions in the organisation of the different levels involved \cite{Horelli01042009}.
Considering the set of projects, research institutions and enterprises that participate to a funding program as a complex dynamical system, it is possible to identify underlying network structures simply defining the edges according to some relations among the components that are of interest for the evaluator.
Once the network is constructed, global and local properties can be evaluated and discussed.

From a data collection perspective, the proposed analysis can profit from current emerging technologies and precise guidelines of European governmental institutions to support initiatives such as \textit{Smart Cities \& Communities}\footnote{\url{http://ec.europa.eu/eip/smartcities/}} and their co-related action goals (\textit{Urban and Citizen App}, \textit{e-Government}, \textit{e-Democracy} and so on). 
All this initiatives have produced a large number of freely available datasets containing information, collected by national governments, which third parties are encouraged to use for their scope, analyse and republish as they wish, without restrictions from any copyright.
Recently, Open Government Data (OGD) is emerging as a major movement in knowledge sharing. 
It promotes transparency and accountability, enables collaboration among stakeholders, encourages novel socio-economic activities and growing of the so-called network economy. 
Starting from the idea that without sharing information it is not possible to establish a culture of collaboration and participation among the relevant stakeholders, 
the Linked Open Data (LOD) \cite{bizer2009linked} Movement, which provides existing data in a machine-readable format, has gained large importance over the last years. 
From a such perspective, LOD facilitates innovation and knowledge creation from interlinked data, but it also introduces a level of complexity for information management and integration. 
Considering a good trade-off between data expressiveness and computational cost for data analysis, we have selected only Open Data repository without linked data and RDF\footnote{\footnotesize Resource Description Framework - \url{http://www.w3.org/standards/techs/rdf\#w3c\_all}} triples. 
Despite the main aim of such movement of reaching the largest possible portion of users, our investigation has outlined that such datasets are usually of heterogeneous quality and size, and that their analysis requires efforts in a pre-processing phase composed of typical ETL (Extract-Transform-Load) \cite{lenzerini2003fundamentals} and data cleaning procedures. 
It is worth mentioning that problems are commonly encountered while using network analysis for evaluation, like the concern about anonymity of non-aggregated data (and eventual anonymisation), or the fact that making results public usually interferes with the structure of the network itself \cite{Penuel01122006}.
These kind of problems are mitigated by using Open Data, since they are public ``by construction''. 

The paper is organised as follows: in \hyperref[sec:methodology]{Sec.~\ref*{sec:methodology}} we introduce the steps composing the schema of the overall analysis process.
In \hyperref[sec:dataset]{Sec.~\ref*{sec:dataset}} we describe the structure of the open data repository of the 2007-2013 Italian Program  \emph{Programma Operativo Nazionale ``Ricerca e Competitivit\`a''} (PON R\&C), in order to keep the paper self-contained, and introduce the data model for network analysis; in \hyperref[sec:network_analysis]{Sec.~\ref*{sec:network_analysis}} we present features and properties of the analysed network. 
In order to better discuss the experimental results, we distinguish among local properties, global properties and community structure. 
\hyperref[sec:conclusions]{Conclusions and perspectives} close the paper.

\section{Methodology}\label{sec:methodology}

\begin{figure}
	\begin{center}
		\includegraphics[scale=0.5]{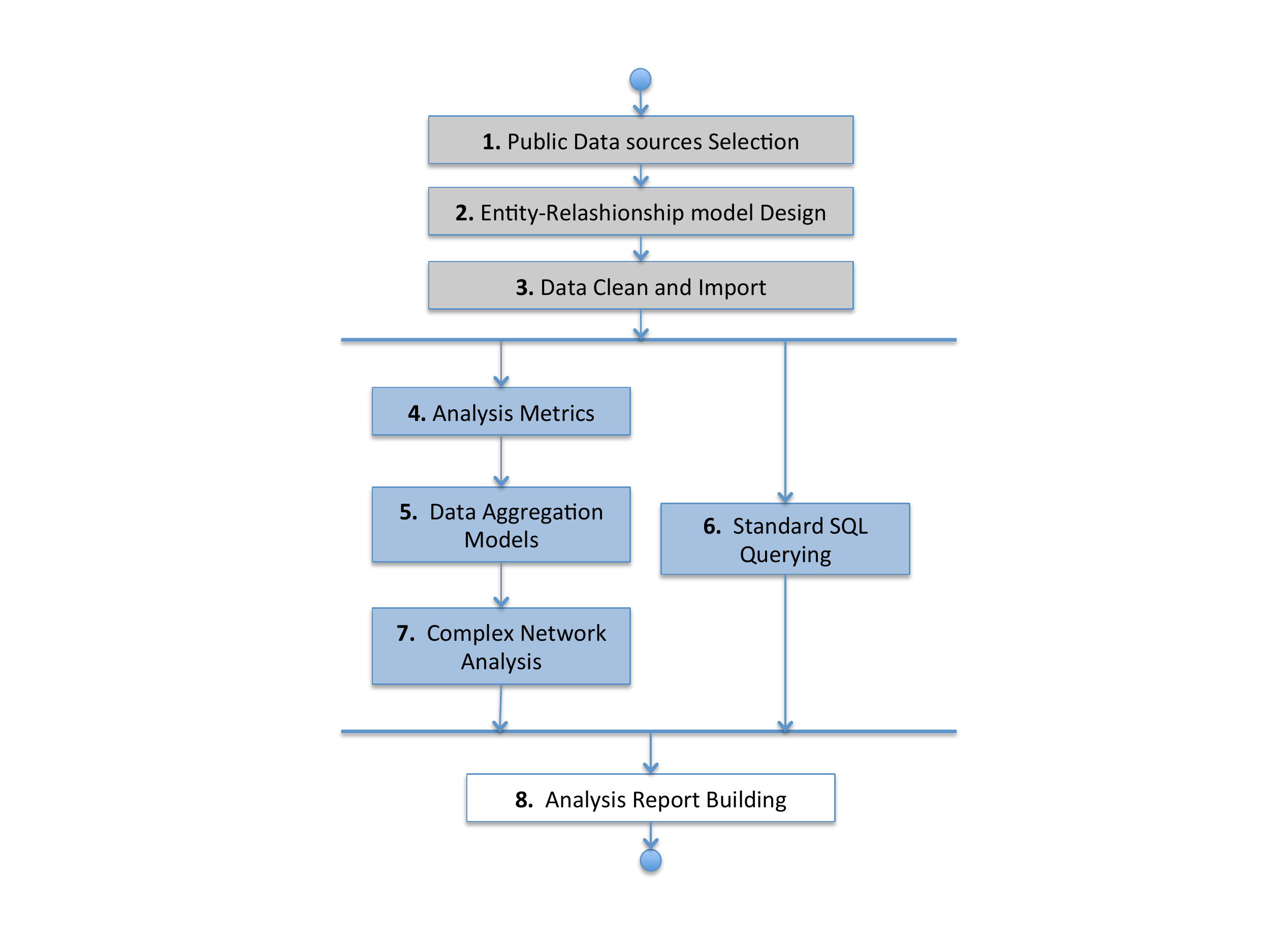}
		\caption{\small Data-flow schema of the overall analysis process}
		\label{fig:dataflow}
	\end{center}
\end{figure}

The approach followed in this paper consists of two main steps shown in \hyperref[fig:dataflow]{Fig.~\ref*{fig:dataflow}}:
\begin{itemize}
	\item processing of data sources (grey blocks)
	\item finding analysis models and metrics (blue blocks)
\end{itemize}
The first blocks of \hyperref[fig:dataflow]{Fig.~\ref*{fig:dataflow}} involve the transformation of a general purpose dataset to an analysis-specific one through the implementation of a data model.
The newly obtained dataset is then used for network analysis.
The graph is constructed identifying the nodes and the properties that define if two nodes are linked or not (in our case: being partner within the same project).
Global and local properties are extracted, in order to produce a qualitative and quantitative description of the structure of the network of relationships generated by the program under examination.
As represented in \hyperref[fig:dataflow]{Fig.~\ref*{fig:dataflow}}, the overall process of our studies ends when a report summarising the analysis and its outcomes is produced.
Descriptions of the numerical outcomes in terms of social/economical effects are given, in order to provide the evaluator with a useful tool for her/his purposes.
We report on the SQL-based \cite{abiteboul1995foundations} modelling approach, which allows to translate a given dataset in Open Data format into the reference set of analysis models (relational tables), and on the selected metrics relevant for executing an effective network analysis. 
It is worth underlining that we have adopted well-known relational tables to store data in order to ensure integrity of our knowledge base and provide significant results.
Our current implementation of data models exploits the open source object-oriented PostgreSQL 9.3 DBMS.
For network analysis we have used the Wolfram Mathematica software and the \texttt{R} programming language.

\section{PON R\&C: From datasets to data models}\label{sec:dataset}

In this Section, the data-driven steps shown in blocks 2--6 of \hyperref[fig:dataflow]{Fig.~\ref*{fig:dataflow}} are described.
As mentioned above, we have selected Open Data about the PON R\&C funding program, publicly available at URL \url{http://www.ponrec.it/open-data/}. 
The program, funded with European Structural Funds managed by \emph{Ministero dell'Istruzione, Universit\`a e Ricerca} (MIUR, Ministry of Instruction, University and Research) and \emph{Ministero dello Sviluppo Economico} (MiSE, Ministry of Economic Development), involved four underprivileged regions in Southern Italy: Apulia, Campania, Calabria and Sicily. 
The main aim of the program consists in promoting socio-economic growth by supporting research and innovation activities, improving  quality of life for citizens and competitiveness of small-medium enterprise (SME). 
The main features of PON R\&C can be summarised as follows:  2962 funded projects, for over 3 billion euro, 11 action programs and 8 action areas: \emph{Health-care}, \emph{Nutrition}, \emph{Energy}, \emph{Environment \& Ecology}, \emph{Transportation \& Logistics}, \emph{Cultural Heritage \& Activities}, \emph{Smart Cities}, \emph{Social Innovation}. 
The group of all the partners involved in each funded project is called \emph{Temporary Scope Association} (TSA) and plays a fundamental role for our network analysis.
The downloaded repository has 3 LOD stars\footnote{\footnotesize The path from Open Data to Linked Open Data was firstly introduced by Sir Tim Berners-Lee in the 5 Stars Model at the Gov 2.0 Expo in Washington DC in 2010, where costs and benefits for both publishers and consumers of LOD are explained.} \cite{janowicz2014five}, is updated at `2014-06-17', and is composed of 3 datasets (files):
\begin{itemize}
	\item \textbf{Projects} -- 10104 tuples with 52 attributes describing project information about program references, activities, textual description of project scope and objectives, details about partners and so on;
	\item \textbf{Locations} -- 11390 tuples with 8 attributes describing details about geographical localisation of project partners; 
	\item \textbf{Budgets} -- 5670 tuples with 13 attributes describing details about amount and state of project funding
\end{itemize}
and one metadata file describing structure and meaning of each the previous files, according to the Open Data standard.
\hyperref[tab:tables]{Tab.~\ref*{tab:tables}} shows a sketch of \emph{Projects}, \emph{Locations} and \emph{Budgets} files, representing information useful for the following discussion in form of couples ``attribute/value''.
\begin{table}[ttt!]
	\begin{center}
		\begin{center}
			\tiny{
				\begin{tabular}{cccc}
					\begin{tabular}{|c|c|c|c|c|c|c|c|} \hline
						\multicolumn{8}{|c|}{\rule [-0.2pt]{0pt}{10pt}\textsc{Project}} \\ \hline
						\rule [-0.2pt]{0pt}{10pt}\texttt{UPC} & \texttt{title} & \texttt{smart\_cities} & \texttt{social\_innovation} & \ldots & healthcare & \texttt{FC} & \texttt{name} \\ \hline\hline
						\rule [-0.2pt]{0pt}{10pt}PON04a2\_A & PRISMA & 1 & 0 & \ldots & 1 & 84001850589 & INFN \\ \hline
                                                     \rule [-0.2pt]{0pt}{10pt}PON04a2\_A & PRISMA & 1 & 0 & \ldots & 1 & 84001850589 & INFN  \\ \hline
						\rule [-0.2pt]{0pt}{10pt}PON04a2\_A & PRISMA & 1 & 0 & \ldots & 1 & 80002170720 & UNIBA \\ \hline
						\ldots &  \ldots &  \ldots &  \ldots &  \ldots &  \ldots &  \ldots &  \ldots \\ \hline
					\end{tabular}
				\end{tabular}
				}
		\end{center}
		\begin{center}
			\tiny{
				\begin{tabular}{cccc}
					\begin{tabular}{|c|c|c|c|c|c|} \hline
						\multicolumn{6}{|c|}{\rule [-0.2pt]{0pt}{10pt}\textsc{Location}} \\ \hline
						\rule [-0.2pt]{0pt}{10pt}\texttt{UPC} & \texttt{FC} & \texttt{name} & \texttt{kind} & \texttt{region} & \ldots \\ \hline\hline
						\rule [-0.2pt]{0pt}{10pt}PON04a2\_A & 84001850589 & I.N.F.N. & PRI & Apulia & \ldots \\ \hline
						\rule [-0.2pt]{0pt}{10pt}PON04a2\_A & 80002170720 & UNIBA & University & Apulia & \ldots \\ \hline
						\ldots & \ldots & \ldots & \ldots & \ldots & \ldots\\\hline
					\end{tabular}
				\end{tabular}
				}
		\end{center}
		\begin{center}
			\tiny{
				\begin{tabular}{cccc}
					\begin{tabular}{|c|c|c|c|c|c|} \hline
						\multicolumn{6}{|c|}{\rule [-0.2pt]{0pt}{10pt}\textsc{Budget}} \\ \hline
						\rule [-0.2pt]{0pt}{10pt}\texttt{UPC} & \texttt{FC} & \texttt{name} & \texttt{total\_cost} & \texttt{total\_funded} & \ldots \\\hline\hline
						\rule [-0.2pt]{0pt}{10pt}PON04a2\_A & 84001850589 & INFN - Apulia & 2231915.7 & 1785532.57 & \ldots \\ \hline
						\rule [-0.2pt]{0pt}{10pt}PON04a2\_A & 80002170720 & University of Bari & 2052539 & 1642031.2 & \ldots \\ \hline
						\ldots & \ldots & \ldots & \ldots & \ldots & \ldots \\ \hline
					\end{tabular}
				\end{tabular}
				}
		\end{center}
		\caption{Sketch of the structure of original files from PON R\&C. Example rows report some features of project titled PRISMA ($UPC=PON04a2\_A$). Note how: 1) \emph{INFN} appears three times with three different values of the \texttt{name} attribute: \emph{INFN}, \emph{I.N.F.N.} and \emph{INFN - Apulia}; 2) in \textsc{Project} table INFN has two duplicate rows.} 
		\label{tab:tables}
	\end{center}
\end{table}
It is important to underline that the approach taken here is somehow different from the ones usually adopted when performing network analysis in other fields.
We have adopted intensive database techniques, while, for example, the treatment of authors of scientific papers with the same name in a social network of scientific collaboration is done automatically by computer algorithms, and errors like the correct identification of the same author represented by two different names (\emph{e.g.} \emph{J.~Smith} and \emph{John~Smith}) are not solved, but just treated statistically \cite{PhysRevE.64.016131,PhysRevE.64.016132}.
We think that, while this is perfectly reasonable in that case, when studying a productive system like the one we are interested in, all the actors must be correctly identified, and in general errors like these should be reduced to the minimum, in order for the analysis to be reliable and really usable by policy makers.

\begin{figure}
	\begin{center}
		\includegraphics[scale=0.5]{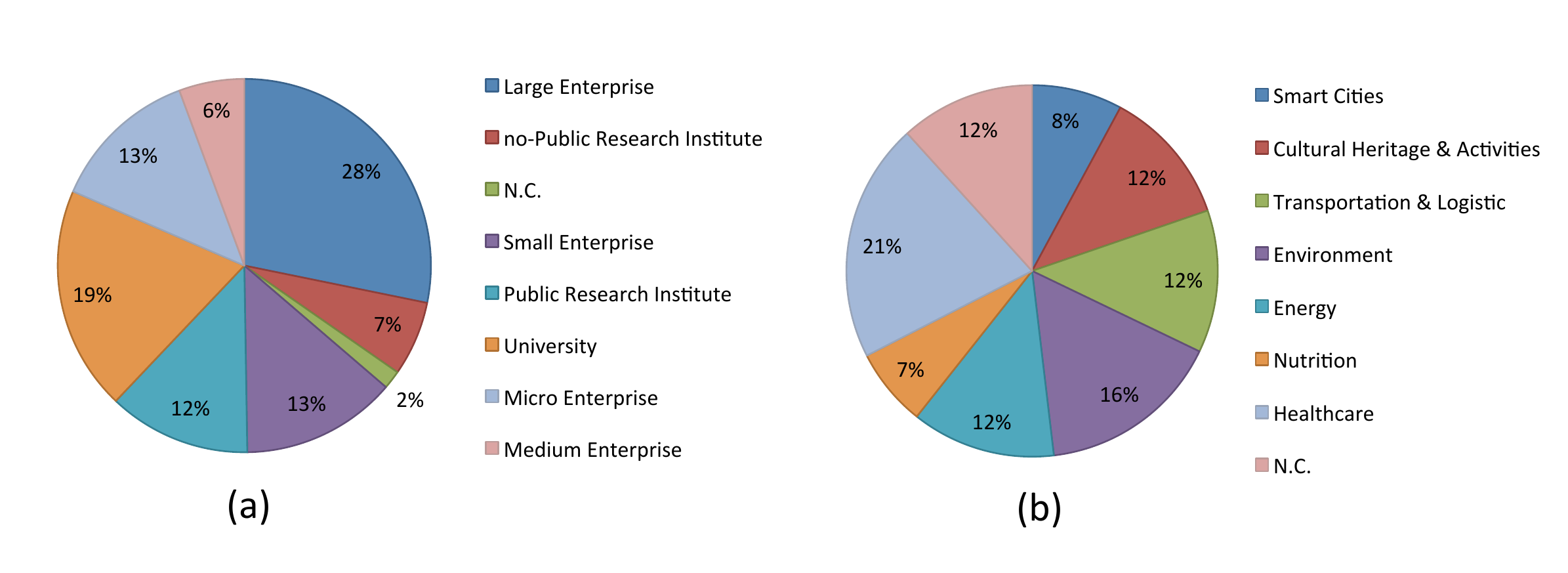}
		\caption{\small Distribution of fundings for 300 projects within the PON~R\&C program. Part \textbf{(a)} represents cost distribution among different kinds of partners, expressed in percentage w.r.t. the total cost; part \textbf{(b)} shows the cost distribution among the action areas mentioned above (exception made for the empty \emph{Social Innovation} one).}
		\label{fig:cost_a_b}
	\end{center}
\end{figure}

\begin{figure}
	\begin{center}
		\includegraphics[scale=0.5]{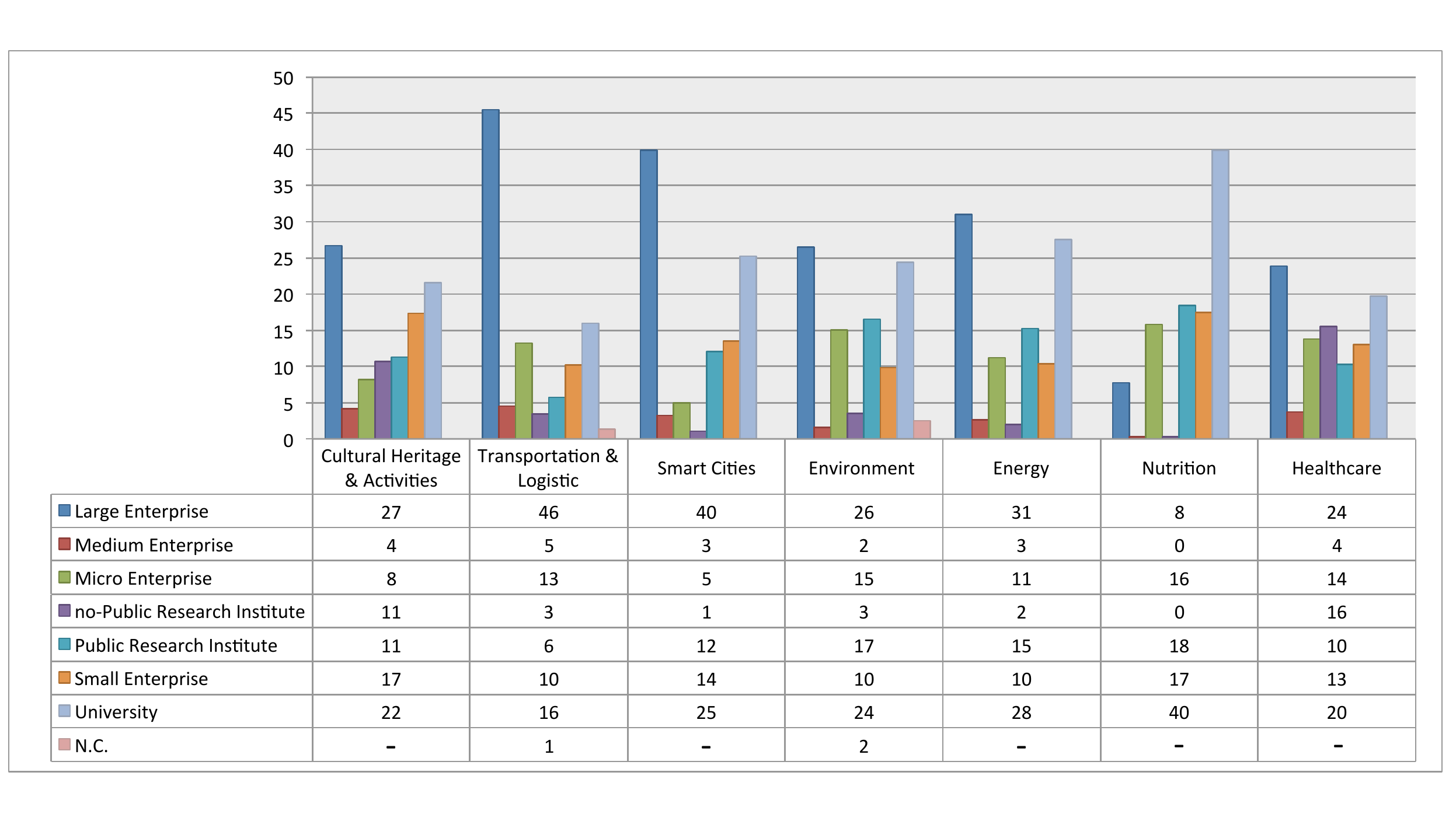}
		\caption{\small Distribution of fundings for each kind of partner, for each action area.}
		\label{fig:cost}
	\end{center}
\end{figure}

In order to accomplish this task, we have defined a proper database model able to store all the projects data. 
First, we have created one table for each of the previous datasets, exploiting the following keys to link tables: \texttt{UPC} attribute is the unique identifier of projects (Unified Project Code) and \texttt{FC} is the unique identifier of partners (Fiscal Code). 
In order to improve dataset quality we have solved textual description encoding and numerical value format.
Moreover, we have overcome name mismatching by inserting a unique label for each partner. 
Such label represents a convenient choice among the multiple names associated to the same fiscal code (FC) in the original datases\footnote{\footnotesize see \emph{e.g.} the \emph{name} attribute in \hyperref[tab:tables]{Tab.~\ref*{tab:tables}}} (\emph{e.g.}, between ``\texttt{I.N.F.N.}'' and ``\texttt{INFN - Apulia}'' associated to the same FC ``\texttt{84001850589}'', we have chosen ``\texttt{INFN}'' as label).
We underline that this data cleaning step is a key aspect in evaluations based on network analysis, in which results are sensitive to lacking data, and it is not possible to sample the population to extract useful information \cite{olejniczak2012evaluating}.
We are aware of the fact that such a problem could be much more evident in very large databases (\emph{i.e.} the ones containing millions or more tuples, rather than thousands, like in the case under examination here), and we think the only viable solution is pushing institutions towards producing better open data.
After the procedure described above, we have obtained a normalised database designing a many-to-many relation among \textsc{Project} and \textsc{Partner} tables and deleting duplicated and bad-formed tuples.
Exploiting SQL standard queries, from our database we have selected 300 projects with at least 2 partners (thus suitable for network analysis). 
Those involve 769 distinct partners, for a total cost of the projects of 2500 \emph{M}euro (around 78\% of the total cost of PON R\&C projects), divided into Universities (33), Public Research Institutes (21), non-Public Research Institutes (44), Micro Enterprises (203), Small Enterprises (232), Medium Enterprises (58), Large Enterprises (163). 
It is significant that $\sim10\%$ of the total number of funded projects, the ones involving a network of relations, represents $\sim78\%$ of the total budget.
This is an indication of the importance of relations in the Italian productive system.
We note that 24 partners, we name N.C. partners, are not classified (in the original datasets) and that the \emph{Social Innovation} action area does not include any project.  
In \hyperref[fig:cost_a_b]{Fig.~\ref*{fig:cost_a_b}}, the distribution of fundings for the selected 300 projects is shown.
\hyperref[fig:cost_a_b]{Fig.~\ref*{fig:cost_a_b}} ~(part \textbf{(a)}) represents cost distribution among different kinds of partners, expressed in percentage w.r.t. the cost of all selected projects, divided for kind of partners and \hyperref[fig:]{Fig.~\ref*{fig:cost_a_b}} ~(part \textbf{(b)}) the cost distribution among action areas above mentioned (except the empty \emph{Social Innovation} area). 
As a general overview,  \hyperref[fig:]{Fig.~\ref*{fig:cost}} shows the distribution of fundings for each kind of partner, for each action area. 
In the calculation of such cost distribution, we have considered the attribute \texttt{total\_cost}, among different ones concerning budgets, since it is the only one having a \texttt{NOT NULL} value for each tuple in the original dataset. 
Starting from concepts underlying the TSA definition, we have built a set of SQL stored procedures performed on our database. 
Inputs are tables and attributes representing significant elements for the networks construction and outputs are the ad-hoc generated views with aggregated and derived data. 
The main generated data views are the following:
\begin{itemize}
	\item \textbf{Partner-to-Partner} -- the distinct couples of partners involved in the same project;
	\item \textbf{Project-to-Project} 	-- the distinct couples of projects having at least one partner in common, together with the calculated number of shared distinct partners;
	\item \textbf{Project-to-Partner}	-- the set of distinct involved partners  for each project; 
	\item \textbf{Partner-to-Funding} --  the funding, for each beneficiary, calculated considering all the PON R\&C projects it is involved in; 
	\item \textbf{Beneficiary-to-Beneficiary} -- the distinct couples of beneficiaries having at least one project in common, together with the calculated number of shared distinct projects.
\end{itemize}	

\section{Network Analysis}\label{sec:network_analysis}

\begin{figure}
	\begin{center}
		\includegraphics[scale=0.18]{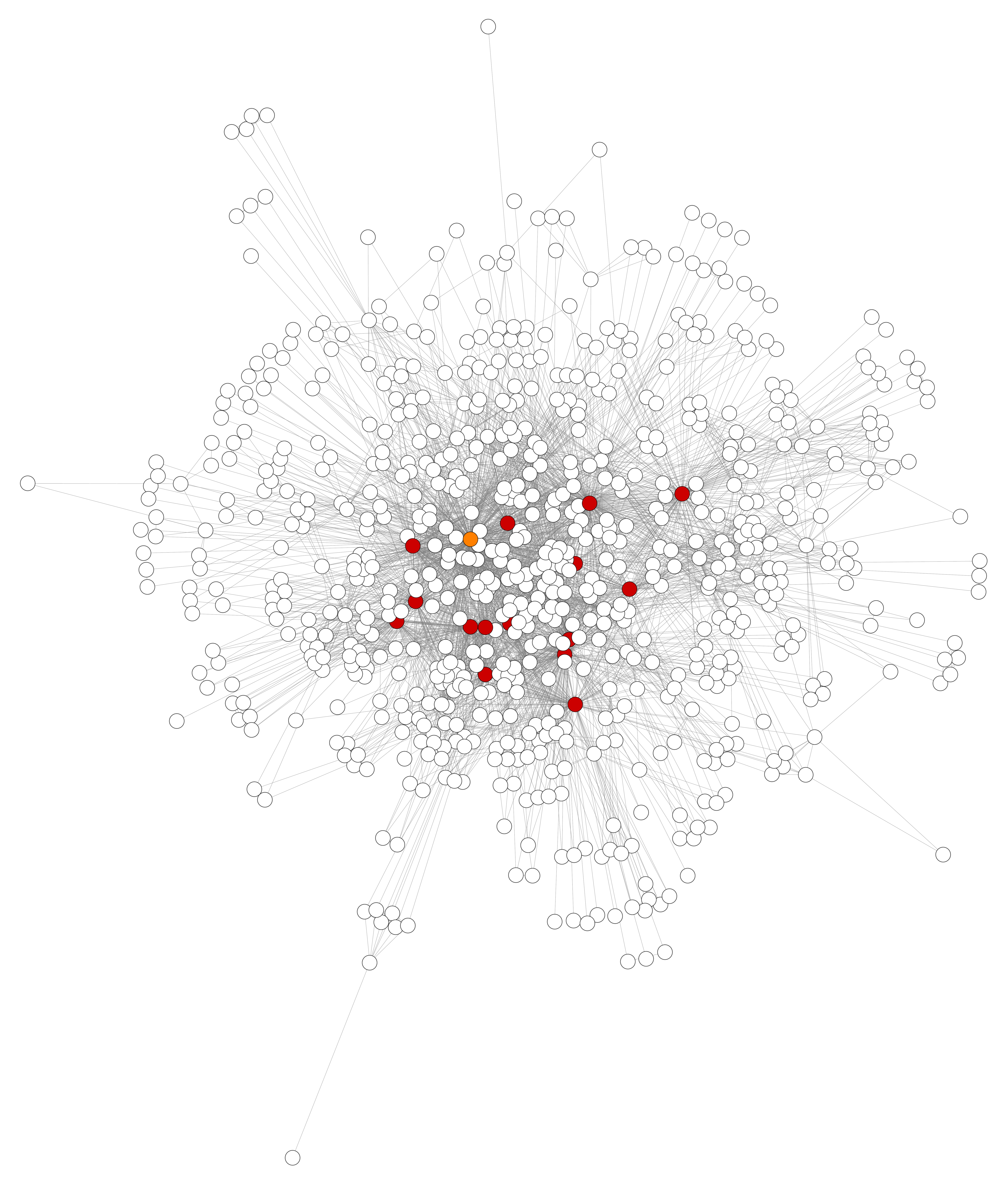}
		\caption{\small Network structure of the Italian PON R\&C funding program. Each vertex is a university, research institution or enterprise funded by the program; two vertices are connected if they are part of a TSA for at least one project. Only the principal giant component is depicted (other connected components have less than six vertices each). The set of nodes constituting the centre of the (giant component of the) PON R\&C network is highlighted in red, while the main hub is depicted in orange. The centre include public and private research institutions, all the major Universities involved in the program, and also some large private enterprises.}
		\label{fig:pon_rec_network}
	\end{center}
\end{figure}

This Section contains a description of the activities described by points 7--8 of \hyperref[fig:dataflow]{Fig.~\ref*{fig:dataflow}}.
The network analysed here is an \emph{affiliation network} \cite{PhysRevE.64.016131,PhysRevE.64.016132}, constructed in such a way that every university, research institution or enterprise that has been funded by the program is a vertex of the graph and there is an edge between two vertices if the corresponding participants are part of at least one TSA, for at least one funded project.
In this way, the graph is the union of complete, undirected, unweighted graphs, each representing a TSA, in which every node is connected to all the others.
The network structure is due to vertices participating to more than one projects, in more than one TSA.
In our analysis we have not considered vertices that have been funded without participating in any TSA. 
The resulting network is shown in \hyperref[fig:pon_rec_network]{Fig.~\ref*{fig:pon_rec_network}}; it has 769 vertices and 4868 edges, and is not connected.
It is made of one giant component, composed of 744 vertices and 4845 edges, and 10 small complete graphs of order 5 (one graph), 3 (two graphs) and 2 (seven graphs).
The graph has been analysed, and several properties \cite{boccaletti06,RevModPhys.74.47} have been extracted to support the  evaluation of the PON R\&C public funding program.
Such properties belong to two main classes: local and global ones.
Local properties are features of single vertices or edges, and, in particular, \emph{centrality coefficients} are evaluated, in order to understand the importance of the nodes within the network.
Global properties involve the network as a whole instead, and are used to describe the full program, independently of the single nodes.

\subsection{Local properties}\label{subsec:local_properties}

Properties of vertices evaluated in the present analysis include \emph{degree centrality}, \emph{betweenness centrality}, \emph{closeness centrality}, \emph{eccentricity}, \emph{eigenvector centrality}, \emph{radiality centrality} and \emph{PageRank centrality}, based on the Google PageRank algorithm \cite{Brin1998107}.

The highest values of all centralities is found in correspondence to public research institutions, like universities and specific research centres.
In particular, the Italian National Research Centre (\href{http://www.cnr.it/}{\emph{CNR}}) shows the best values for all the indicators.
It is worth saying that it is a peculiar vertex of the network, being composed of 104 institutes spread over geographically distributed sites (in all the biggest cities in Italy), and covering a large spectrum of activities in many fields, from pure research, to applied disciplines.
Probably, it would be better to split such vertex and consider each site, or department, separately, but the dataset does not contain such details.
On the contrary, dividing the CNR in many entities would in a certain sense spoil its central nature in the Italian panorama.
Resolving this controversy is interesting, but is over the purposes of the present paper, and is left for a future work, when more detailed Open Data will be available.
Apart from cases like the one described, the central role of public research institution for the network structure is clear from all the centralities.

Degree centralities are discussed in detail in the next \hyperref[subsec:global_properties]{Sec.~\ref*{subsec:global_properties}}, since global properties of the network can be inferred from the distribution of such quantities, despite being them local in nature. 

Betweenness \cite{1977} measures the importance of a node for traffic of information across the network.
Large betweenness centrality of a vertex indicates that many shortest paths between couples of other vertices pass through that node.
The relevance of this quantity for program evaluation stands in the possibility of assessing the role of institutions/enterprises for the eventual aggregation of ``far'' nodes.
For example, a policymaker interested in promoting a program aimed at aggregating and consolidating the productive system of a region should pay attention not to spoil the edge betweenness of the network of relations between the actors involved in the program.

Closeness centrality indicates whether a node is at a short average distance from every other reachable vertex, with higher closeness meaning shorter distance.
A variant is radiality centrality, which gives higher weight to the neighbourhood of the node.
From the social/economical point of view these quantities give indication about how easily an institution/enterprise can connect to all the other members of the network (and, so, of the productive system).
For example, an enterprise with high closeness centrality could be the right promoter for initiatives like the creation of technological districts, associations or lobbies. 
Exploiting the information contained in this quantity, a policymaker could more easily head the productive system in the desired direction with focused regulatory interventions.

Eccentricity is the maximum value of the distances between a node and any other node in the network.
It gives an idea of how central a vertex is within the network, with smallest values corresponding to more central nodes.

High eigenvector centrality is assigned to vertices that are connected to many other well-connected vertices.
It can be used to identify the best way to spread a trend within the productive system represented by the network.
A variant of eigenvector centrality is PageRank centrality, which is a way of measuring the importance of a node within a graph.
The original algorithm was created by Larry Page and Sergey Brin in 1986 at Stanford University \cite{page2001method,page2011annotating,page2014scoring} and is widely used by Google to measure the importance of website pages.
The algorithm used here is given by the solutions of:
\begin{equation}
	\mathbf{r}=\alpha\,{\CMcal A}^{\intercal}{\CMcal H}\,\mathbf{r}\,\,,
\end{equation}
where $\mathbf{r}$ is the vector of the PageRank centralities for each node, ${\CMcal A}$ is the adjacency matrix of the graph, ${\CMcal H}$ is the diagonal matrix consisting of $1/\max\{1,d_{i}\}$, $d_{i}$ being the degree of the \emph{i}$^{th}$ vertex, and $\alpha$ is the damping factor, an empirical parameter\footnote{representing the probability that a traveler randomly navigating the network continues doing it at a given point} usually set to $\alpha=0.85$ \cite{Brin1998107}.
This is an example of how fruitful the application of social network analysis can be to the evaluation of the effects of public funding, also by means of well known and successful algorithms, as the PageRank one.

\begin{figure}[h!]
	\centering
	\subfigure[Degree Centrality]{
		\includegraphics[scale=0.37]{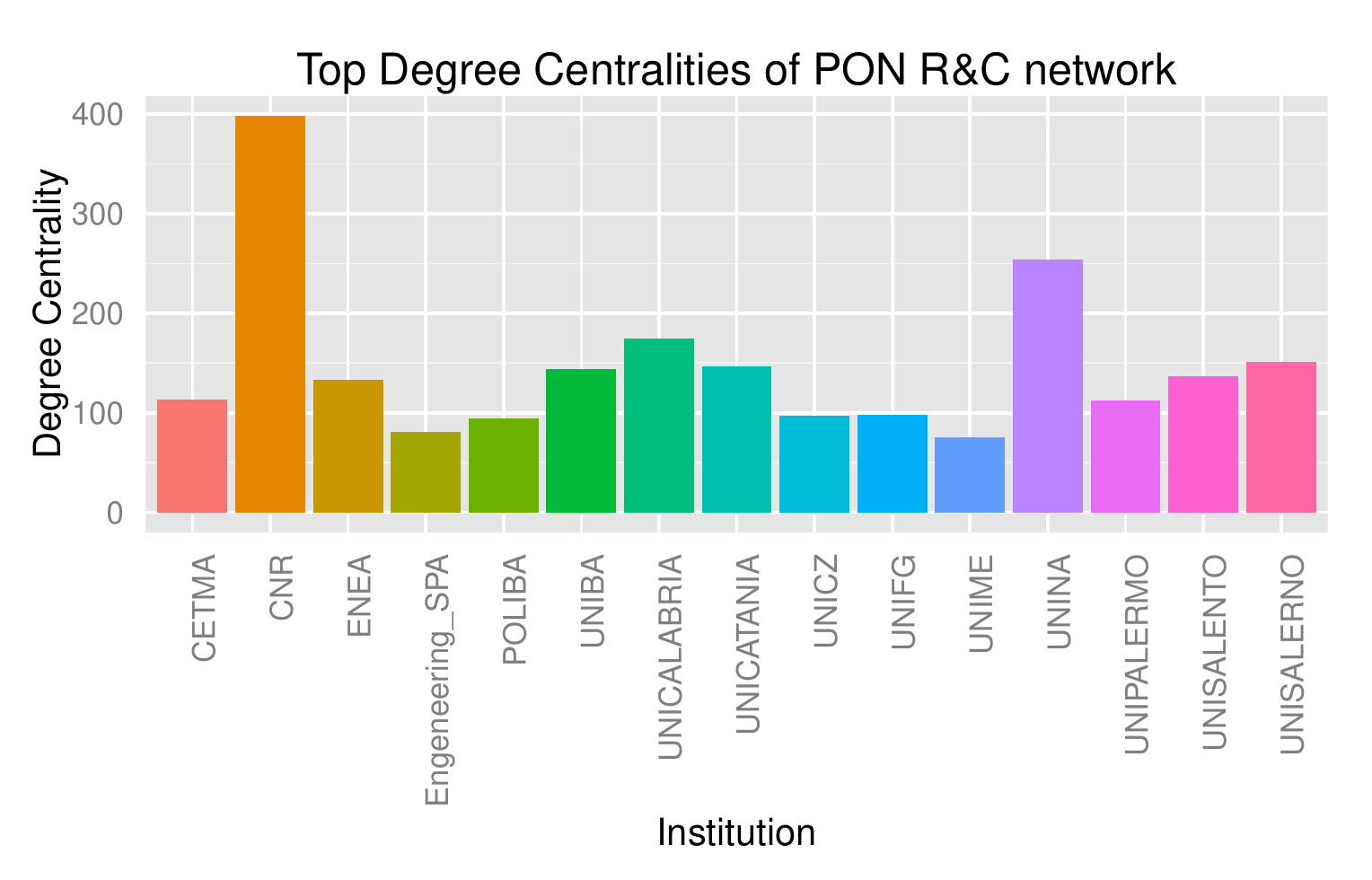}
		\label{fig:vertex_centralities_degree}
	}
	\subfigure[Betweenness Centrality]{
		\includegraphics[scale=0.37]{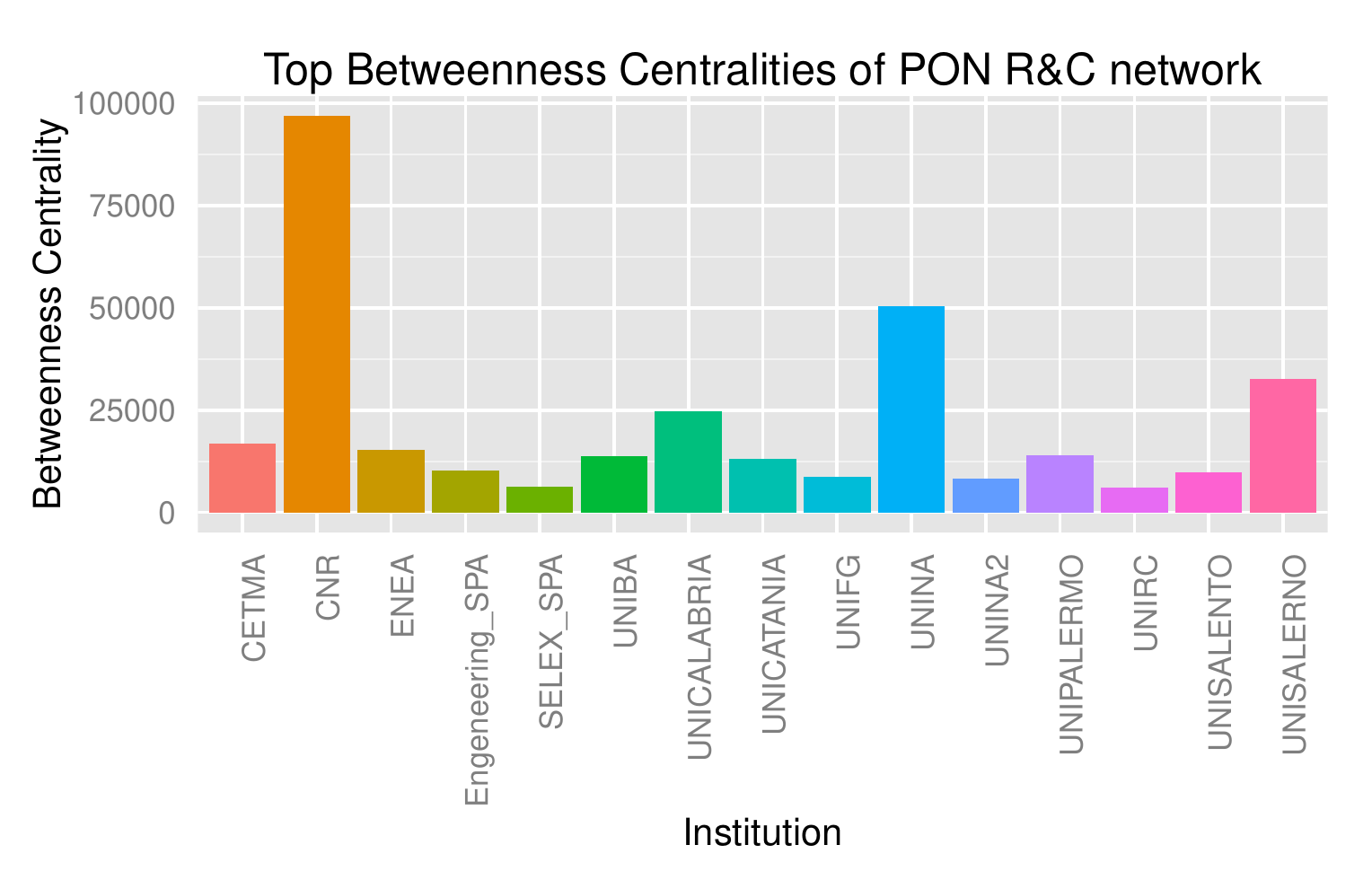}
		\label{fig:vertex_centralities_betweenness}
	}
	\subfigure[Closeness Centrality]{
		\includegraphics[scale=0.37]{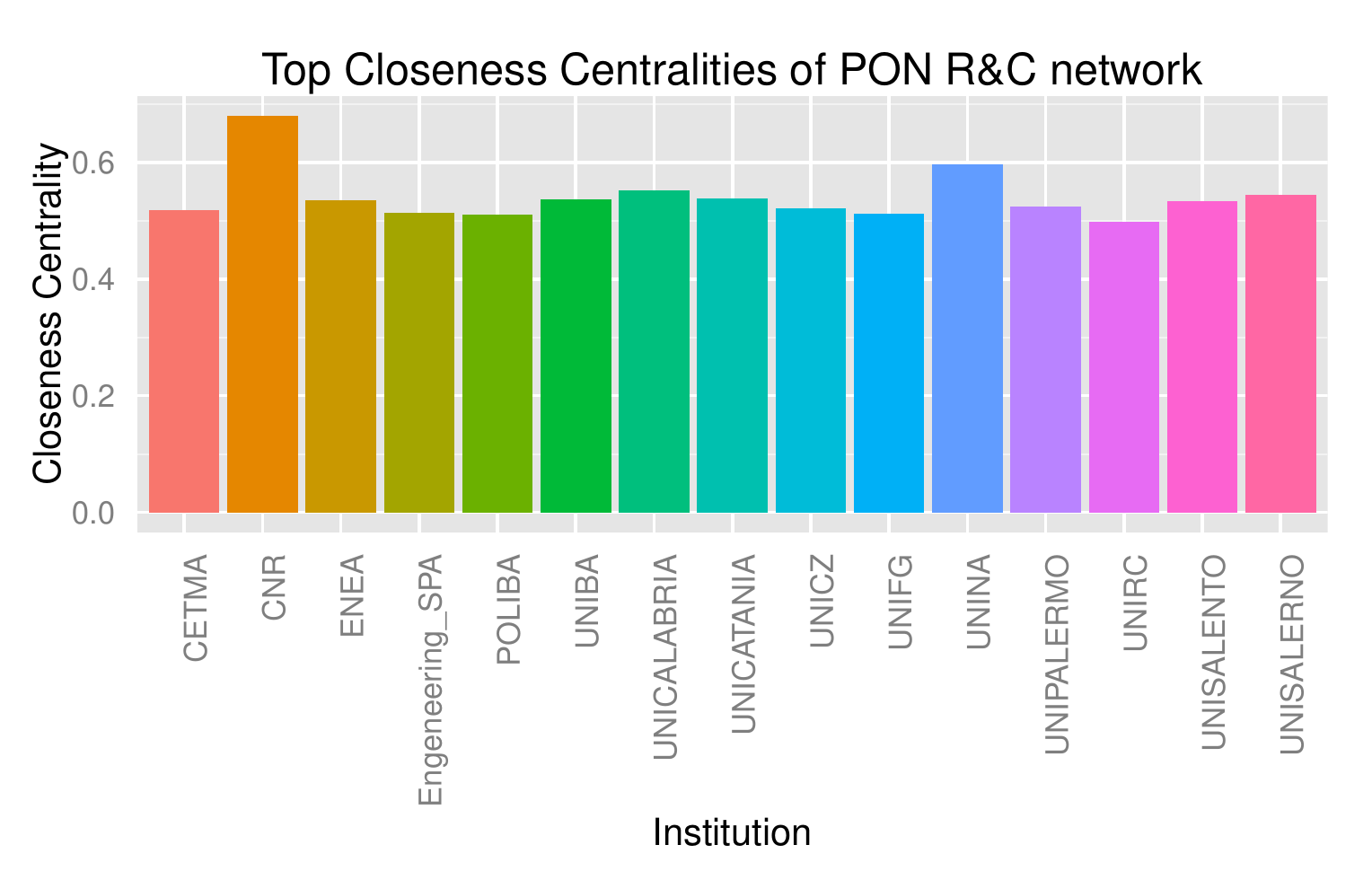}
		\label{fig:vertex_centralities_closeness}
	}
	\subfigure[Eigenvector Centrality]{
		\includegraphics[scale=0.37]{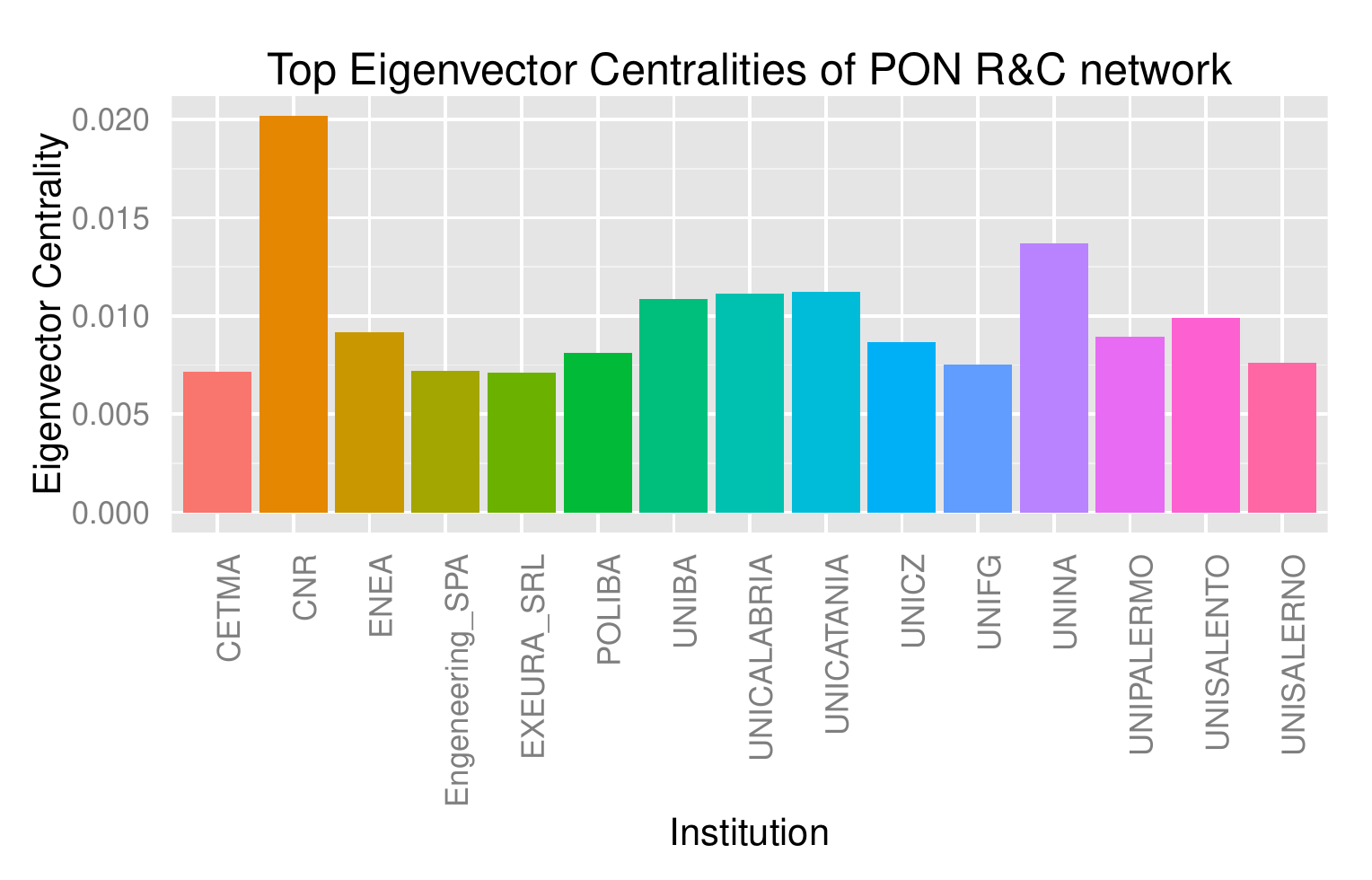}
		\label{fig:vertex_centralities_eigenvector}
	}
	\subfigure[PageRank Centrality $(\alpha=0.85)$]{
		\includegraphics[scale=0.37]{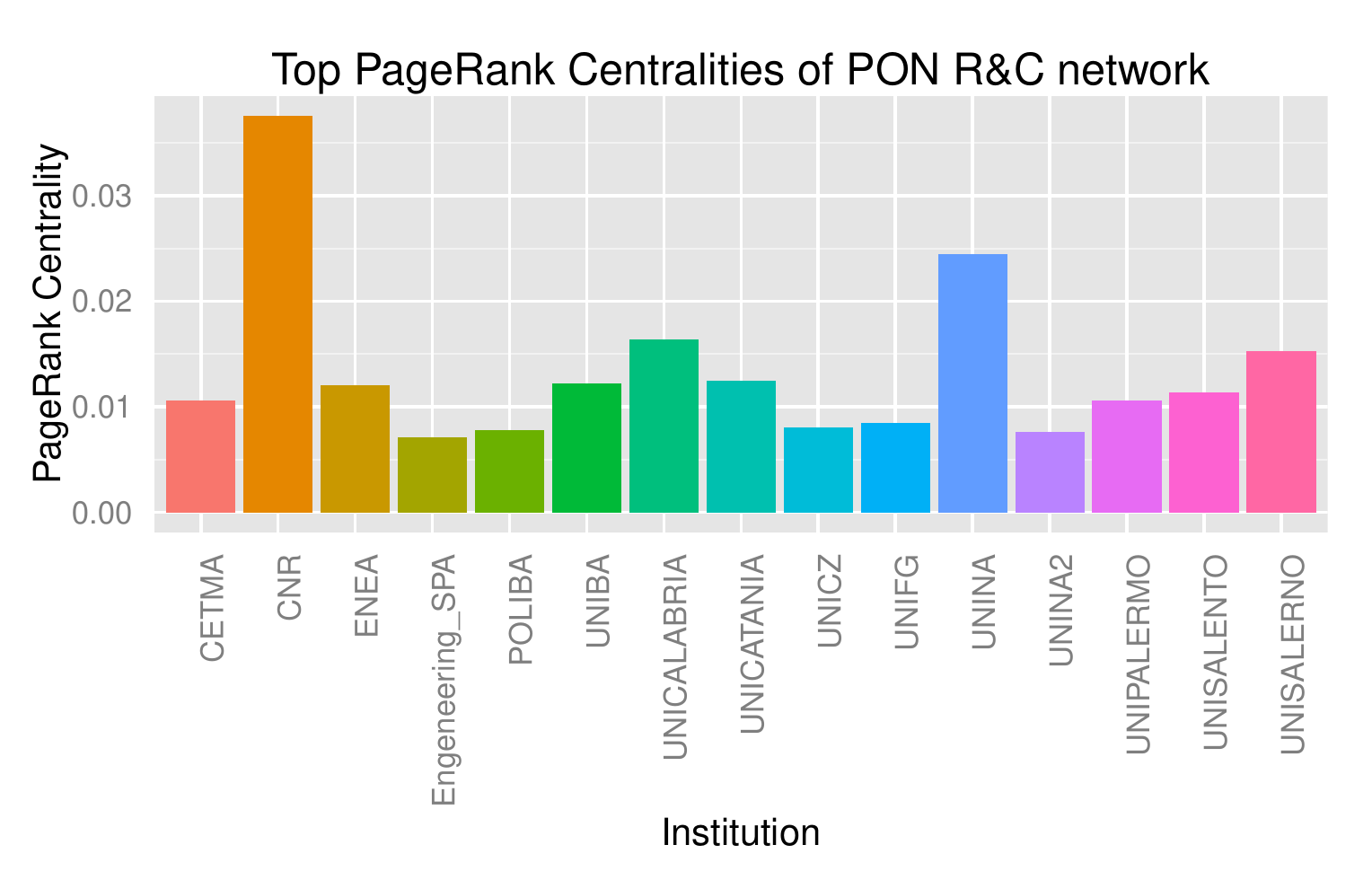}
		\label{fig:vertex_centralities_pagerank}
	}
	\subfigure[Radiality Centrality]{
		\includegraphics[scale=0.37]{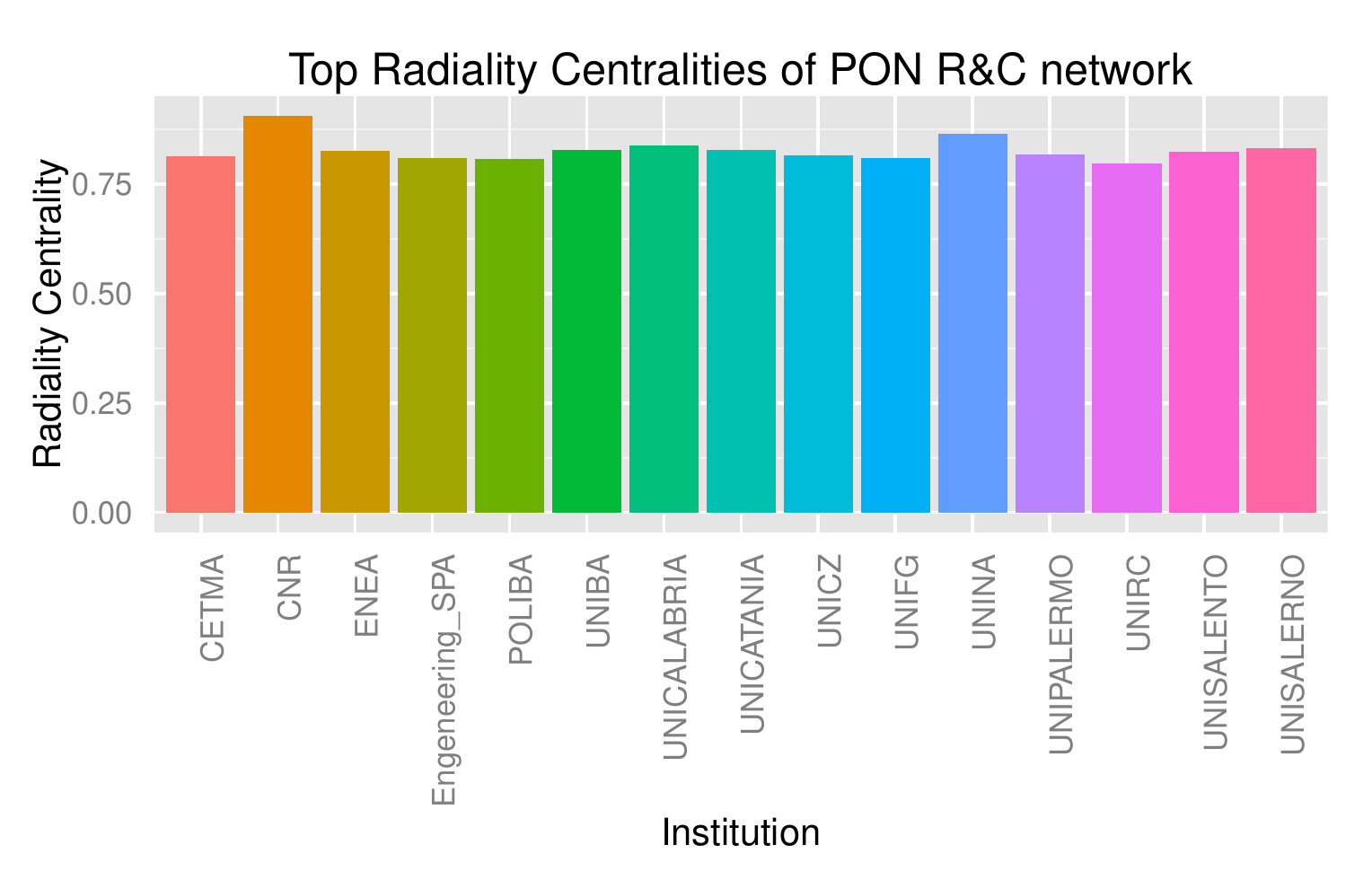}
		\label{fig:vertex_centralities_radiality}
	}
	\caption{\small The fifteen largest values of each vertex centrality for the (giant component of the) PON R\&C network. The highest positions are occupied by public research institutions.}\label{fig:vertex_centralities}
\end{figure}

In \hyperref[fig:vertex_centralities]{Fig.~\ref*{fig:vertex_centralities}} the largest values of all the centralities of the (giant component of the) PON R\&C network are reported.
As stated before, the CNR has the highest values for all the centralities, probably due to its being scattered along the whole country.
Nevertheless, it is important to observe that public research institutions, mainly universities, occupy the top positions for every centrality, while in the lowest positions we find private enterprises, no matter whether they are large or small (except for the Polytechnic of Bari, which is a public university and has a low value of PageRank centrality).
The central role of public research institutions for the network of relations underlying (at least part of) the Italian productive system is clear from \hyperref[fig:vertex_centralities]{Fig.~\ref*{fig:vertex_centralities}}.
Eccentricity is not reported in the figure, since a high number of vertices share the same value of this centrality, meaning that the network is somewhat ``equally spaced''.
For degree and betweenness centralities only the largest values are reported, since the smallest ones are trivial.

\begin{figure}
	\begin{center}
		\includegraphics[scale=0.5]{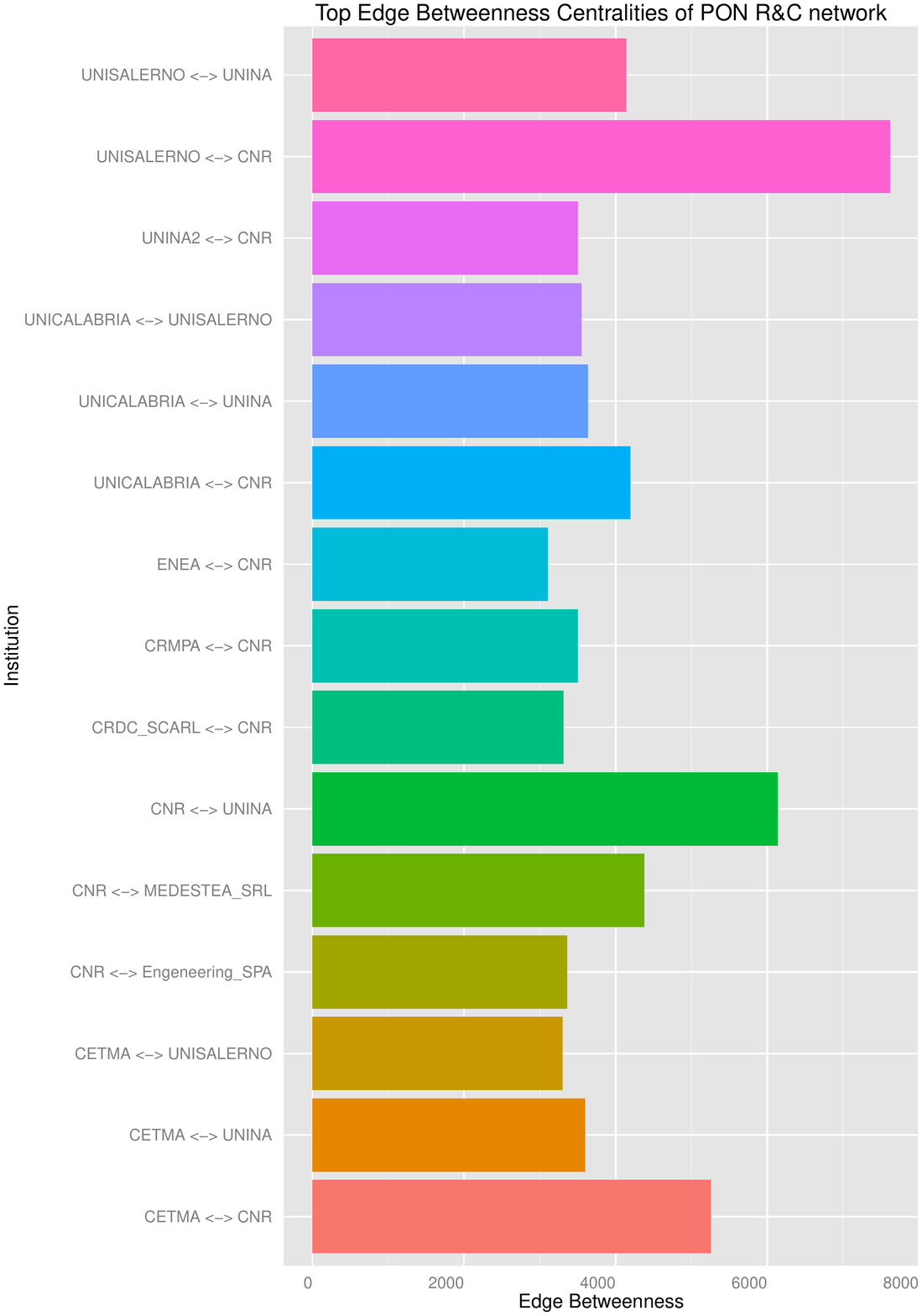}
		\caption{\small The fifteen largest values of edge betweenness centrality for the (giant component of the) PON R\&C network. The highest positions are occupied by links involving public research institutions.}
		\label{fig:edge_betweenness}
	\end{center}
\end{figure}

The last centrality considered here is edge betweenness, which is a property of the edges of the network (rather than vertices), and measures how central a link is for the connections between nodes.
It is measured by counting the number of shortest paths the edge belongs to, and gives a quantitative idea of how much the relation between two institutions/enterprises is important for the ``communication'' between all the actors composing the network.
As the number of edges is much larger than the number of vertices, and since it is necessary to evaluate the shortest path between any couple of nodes, the calculation of such centrality is a resource intensive process.
The largest values of the edge betweenness for the (giant component of the) PON R\&C network are reported in \hyperref[fig:edge_betweenness]{Fig.~\ref*{fig:edge_betweenness}}.
Also in this case, the most important relationships (edges) between the nodes of the network are the ones between public research institutions, while small enterprises give small to no contribution to the geodesics.

\subsection{Global properties}\label{subsec:global_properties}

The first property analysed here is the degree distribution of the vertices, \emph{i.e.} the frequencies of the degree centralities described in the previous \hyperref[subsec:local_properties]{Sec.~\ref*{subsec:local_properties}}.
The importance of such distribution stands in the possibility of inferring from it information about the topology of the graph, and in particular to understand if the network is \emph{scale-free} \cite{Barabasi15101999}.
The property of being scale-free is shared by many real networks, showing power law-shaped degree distributions $P(k)=A\,k^{-\gamma}$, with exponents usually varying in the range $2<\gamma<3$, which have the same form at all scales.

This is of particular interest since power laws are commonly associated with second-order phase transitions in dynamical systems.
Phase transitions in complex networks represent an interesting research field \cite{PhysRevLett.85.5234, PhysRevE.66.036140}, but the graph considered here is static, so no considerations can be made in this respect.
Anyway, this is an interesting perspective for a future work, in which dynamics can be taken into account.

Scale-free networks have an inhomogeneous degree distribution, with many nodes having more connections than the average (\emph{hubs}).
The hubs follow a hierarchy, in which large ones are connected to smaller ones, which are themselves connected to even smaller ones, and so on.
This feature makes the network robust against casual failures, since the removal of a random vertex would not systematically affect the main hubs, and connectedness would not be spoiled.
Hence, scale-free graphs are a desirable result for policymakers interested in generating a solid network of relationships between productive actors on the territory.
Apart from being a strong point for networks, hubs also represent a weakness, since their systematic removal would quickly destroy the network.
The property of being scale-free is an important point to be taken into account for an evaluator, as we will show below, in order to monitor and evaluate the results of funding programs.
Moreover, it suggests to decision makers that effort should be put in promoting funding program which hubs can profit from.

\begin{figure}[h!]
	\centering
	\subfigure[Degree Distribution]{
		\includegraphics[scale=0.38]{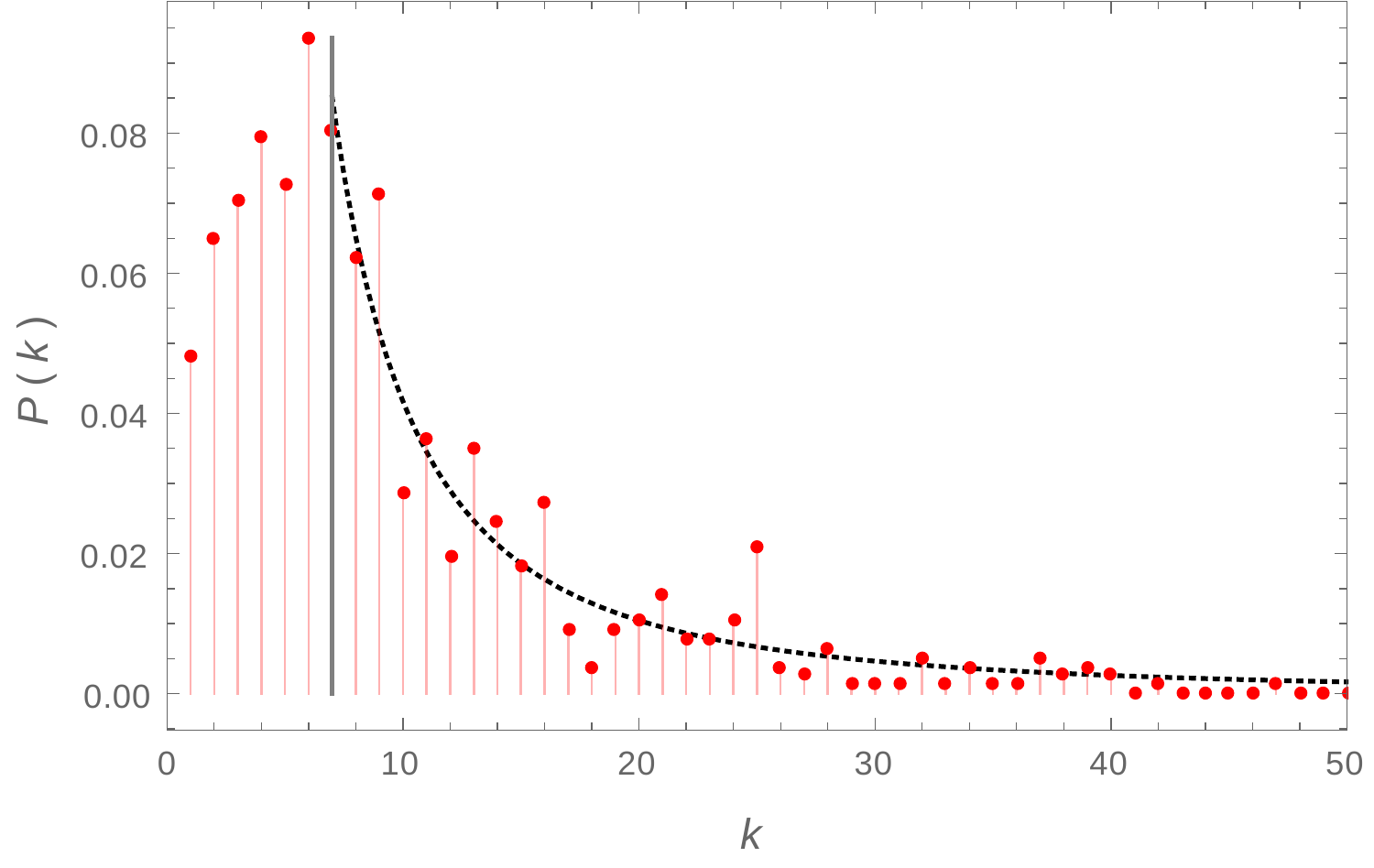}
		\label{fig:pon_rec_degree_distribution_lin}
	}
	\subfigure[Degree Distribution -- Log-Log Plot]{
		\includegraphics[scale=0.38]{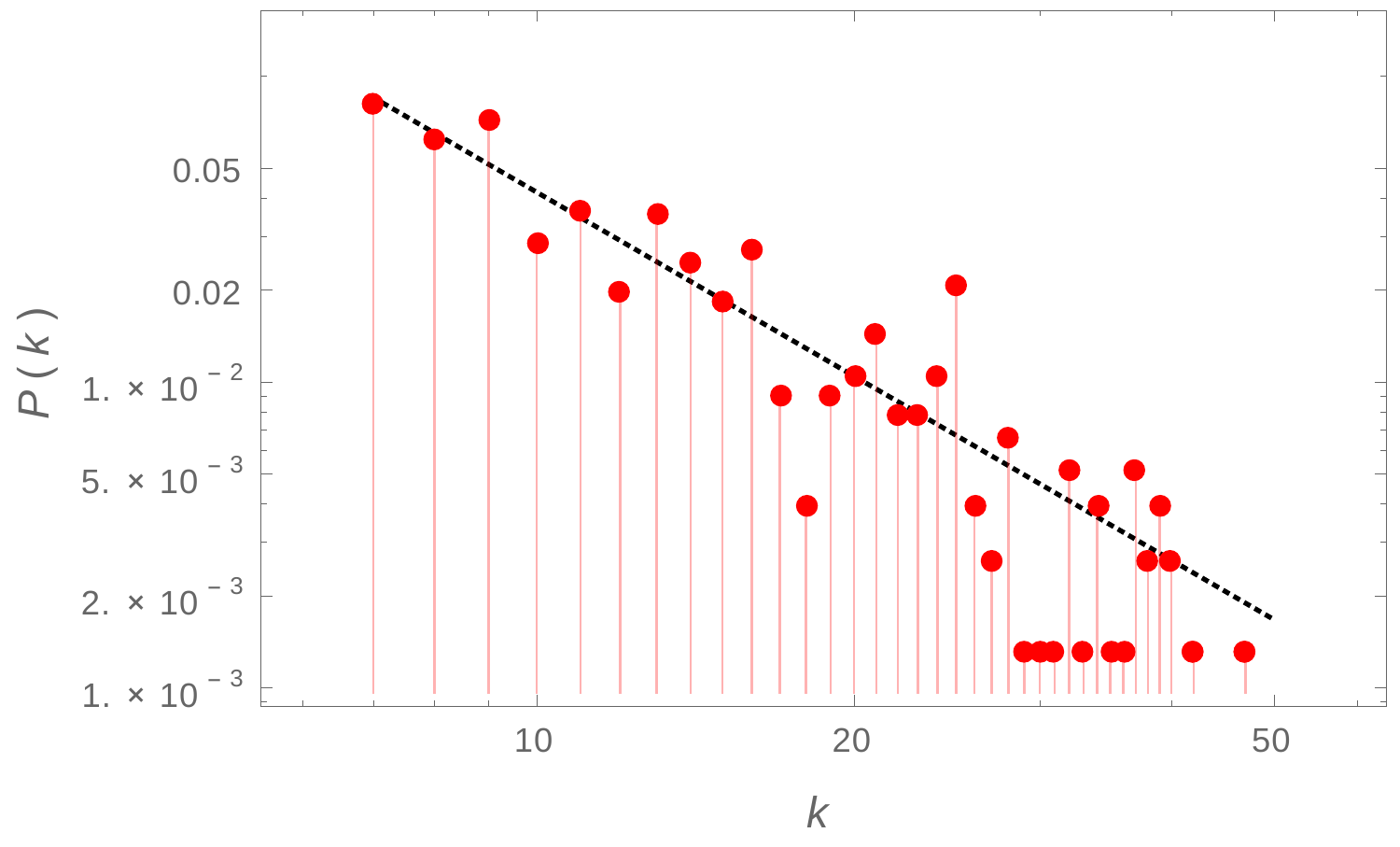}
		\label{fig:pon_rec_degree_distribution_log_log}
	}
	\caption{\small The degree distribution of the network based on the PON~R\&C funding program is shown. The tail is fitted to a power-law function of the form $P(k)=A\,k^{-\gamma}$ with $A=4.156\pm0.375$ and $\gamma=1.998\pm0.040$. The vertical grey line in \hyperref[fig:pon_rec_degree_distribution_lin]{Fig.~\ref*{fig:pon_rec_degree_distribution_lin}} shows the position of the cutoff (corresponding to the upper bound of the median interval) used to fit the tail. Part (b) shows the tail and the fitted line in Log-Log scale.}
 	\label{fig:pon_rec_degree_distribution}
\end{figure}

The degree distribution of the PON~R\&C network is shown in \hyperref[fig:pon_rec_degree_distribution]{Fig.~\ref*{fig:pon_rec_degree_distribution}}.
The tail (starting from the upper bound of the median interval, $m=7$) is fitted very well to a power-law function of the form $P(k)=A\,k^{-\gamma}$ with $A=4.156\pm0.375$ and $\gamma=1.998\pm0.040$.
To obtain the fits, a nonlinear regression based on Newton method \cite{avriel2003nonlinear} has been used.
A comparison with another fit, to an exponential distribution $P(k)=A\,e^{-\gamma k}$ with $A=0.241\pm0.011$ and $\gamma=0.164\pm0.005$, shows that the former fits the distribution slightly better than the latter, with $R_{pow}^{2}=0.935$ and $R_{exp}^{2}=0.929$.
Such a small difference between the values of $R^{2}$ is not a strong indication of the fact that a power-law fits the distribution better than an exponential law, but together with the fact (shown below) that higher moments grow, it is sufficient to assess the power-law nature of the distribution.
In fact, for a power-law distribution with tail of ${\CMcal O}\left(x^{-\nu}\right)$ the moments of order $n\geqslant(\nu-1)$ diverge, and in general higher order moments are larger in size with respect to the lower order ones (this is not true for exponential distributions).
Standing the known difficulties in evaluating the nature of the degree distribution, due to noise coming from the finiteness of the sample (especially from boundary values), the present result is satisfactory in assessing the property of the PON R\&C network of being scale-free.
More refined methods could be used to evaluate the parameter $\gamma$ with higher precision like \emph{e.g.} the Kolmogorov-Smirnov test \cite{Eadie:100342}, but this is outside the purposes of the present work.

The distribution has an expected value $\langle k\rangle=12.661$, mode $M=6$, median $6<m<7$, standard deviation $\sigma=23.781$, skewness $s=8.876$ and kurtosis $\kappa=113.882$ (all the moments are shown in \hyperref[tab:degree_distribution_moments]{Tab.~\ref*{tab:degree_distribution_moments}}).
\begin{table}[h!]
	\begin{center}
		\footnotesize{
		\begin{tabular}{|c|c|}
			\hline\hline		&							\\
			\textbf{Moment}		&	\textbf{Value}				\\
			\hline\hline		&							\\
			Expected value		&	$\langle k\rangle=12.661$	\\
			\hline			&							\\
			Mode			&	$M=6$					\\
			\hline			&							\\
			Median			&	$6<m<7$					\\
			\hline			&							\\
			Standard deviation	&	$\sigma=23.781$			\\
			\hline			&							\\
			Skewness			&	$s=8.876$				\\
			\hline			&							\\
			Kurtosis			&	$\kappa=113.882$			\\
			\hline\hline
		\end{tabular}
		}
		\caption{Moments of the degree distribution of the PON R\&C network.}\label{tab:degree_distribution_moments}
	\end{center}
\end{table}

Once assessed such power-law nature, it is interesting to identify the main hubs.
In the PON~R\&C network considered here, hubs are public research centres, and this represents a strong point for the relationships of the involved productive system.
In fact, it is natural, in the lifecycle of a productive system, that some enterprises rise while others fall, resulting, in the language of networks, in the random removal of vertices described above.
Anyway, as previously stated, the random removal of vertices from a scale-free network does not spoil connectivity, which happens with the systematical removal of the main hubs instead.
In this case, it is unlikely that one of the main hubs, identified here with large public research centres, could disappear, since this would mean \emph{e.g.} the closure of a large public university, a quite rare event.
This picture was partly expected, since in many cases it was mandatory to involve public research institutions in the TSAs.
Nevertheless it still represents a strong indication for a decision maker, suggesting that it is ``safer'' including public research in a future program, since it is the easiest way to keep a solid relationship network within the productive system.

Another way of assessing if a network is scale-free consists in evaluating the distribution of local clustering coefficients, \emph{i.e.} the number of edges connecting the neighbours of each vertex $v$, divided by the number of edges of a complete graph of the same cardinality of the neighbourhood of $v$ \cite{Watts:1998fk}.
The PON R\&C network represents a special case, in which local clustering coefficients are less important, the majority of them being close to 1 by construction. 
In fact, since the graph is a union of complete graphs, it is likely that the neighbourhood of a vertex is fully connected, implying the closeness to one of the local clustering coefficient.
The global clustering coefficient ${\CMcal C}$, \emph{i.e.} the fraction of paths of length two that are closed (over all paths of length two), is much more significant instead, and it takes a small value ${\CMcal C}=0.215$ for the giant component, meaning that the network is not strongly clustered.
From the political and sociological point of view, this is an interesting point, since the network is made by ``scattered'' relationships, despite being composed of ``closed'' TSAs.

Other important features that can guide the policymaker in evaluating the effects of the program or planning future ones are \emph{vertex connectivity} $V_{c}$ and \emph{edge connectivity} $E_{c}$, \emph{i.e.} the smallest number of vertices or edges to be removed in order to disconnect the graph, respectively.
For the case under examination such quantities take value $V_{c}=1$ and $E_{c}=1$, meaning that the removal of a single node or edge can be catastrophic for network connectivity.
Identifying and monitoring such nodes/edges can be very important in case of low values of such parameters, in order to keep the network of relations tightly connected.

Another important property of scale-free networks is that they are \emph{small world} networks \cite{milgram1967swp}.
This means that relatively short paths exist between any two nodes (with respect to the large size of the graph), with an average shortest path length\footnote{\emph{i.e.} \footnotesize the average length of all shortest paths between couples of vertices of the graph.} $L\sim{\CMcal O}(\log N)$, $N$ being the total number of vertices.
This is due to the existence of links between vertices belonging to farther parts of the graph, having the role of connecting them and reducing distances to few hops.
Usually, in scale-free networks such vertices are the hubs and the small-world property is enhanced when $2<\gamma<3$ where $L\sim{\CMcal O}(\log\log N)$ (while it is $L\sim{\CMcal O}(\log N)$ when $\gamma>3$) \cite{Chung10122002}.
For the PON~R\&C network, $\gamma=1.998\pm0.040$, $L=2.532$, $\log N=6.645$ and $\log\log N=1.889$, so the small world property is enhanced, as expected when the vertex distribution follows a power-law with $2<\gamma<3$.
Again, this cannot be considered a smoking gun proving that the network is scale-free, but just another indication in addition to the ones mentioned above.

Other global features of the network are the radius ${\CMcal R}$ and diameter ${\CMcal D}$ of the graph, defined as the minimum and the maximum eccentricity of all vertices, respectively, the eccentricity being the longest shortest path from a source node to every other vertex in the graph.
For the PON~R\&C network ${\CMcal D}=5$ and ${\CMcal R}=3$, meaning that no vertex is more than 5 hops far from any other node, and that the farthest destination is never closer than 3 hops from any source.
From the point of view of program evaluation, this means that that PON~R\&C has been successful in creating (or intersecting) a network of close relationships between the funded actors.
Being interested in promoting such a relationship network while defining the program, these could be good \emph{ex post} indicators of the goodness of the obtained results.

The centre of the graph\footnote{\emph{i.e.} \footnotesize the set of vertices with minimum eccentricity.} is shown in \hyperref[fig:pon_rec_network]{Fig.~\ref*{fig:pon_rec_network}}.
It includes public research centres like CNR (which is also the main hub) and ENEA, all the major Universities involved in the program (Bari, Calabria, Catanzaro, Foggia, Naples, Palermo, Salento, Salerno), private research centres like CETMA, and also some large private enterprises like Avio S.p.A., Engeneering S.p.A., IBM, SELEX S.p.A., and EXEURA S.r.l..
This is a strong indication that the network of funded projects gravitates around large poles involving research centres (public and private), which turn out to have a key role in aggregating entities.
This can also be an explanation for the scale-free property of the graph, since \emph{preferential attachment} is known to be a generating mechanism for this kind of networks \cite{Barabasi1999173,doi:10.1080/00018730110112519,PhysRevLett.85.4629}, in which nodes prefer to link to vertices with high degree.
It is reasonable to imagine that many small actors prefer forming TSAs including large research organisations, which are usually able to get more funds, rather than form TSAs between themselves.
From the point of view of social networks and relationships, it is particularly interesting to study such feature side-by-side with assortativity \cite{PhysRevLett.89.208701}, which indicates whether nodes of the graph tend to connect with their connectivity peers (vertices with similar degree) or not.
In the first case the network is said to be \emph{assortative}, while in the second case it is \emph{anti-assortative}.
This feature is quantitatively measured through the \emph{assortativity coefficient} $r$, whose range is $-1\leqslant r\leqslant 1$, $r=1$ ($-1$) meaning a perfectly (anti-)assortative graph and $r=0$ indicating no particular preference for the majority of the nodes.
In the present network, $r=-0.173$, meaning that the graph is slightly anti-assortative.
This means that the productive system funded by such program has a little tendency not to form \emph{lobbies} among important actors, but to associate strongly connected hubs to smaller and less connected enterprises/institutions.
From the socio-economical point of view, it seems reasonable that small enterprises turn to larger ones or to big research centres to benefit from sharing and collaborations
This is an interesting result, since most social networks show assortative mixing by node degree \cite{PhysRevE.67.026126}, and it also has some implications on the topology of the network.
First, anti-assortative networks are more susceptible to the removal of high-degree nodes (here represented by universities and research centres), which is an indication for the policymaker of the importance that public research has in the productive system, and of the possible disruptive effect of its underestimation.
Second, in anti-assortative networks epidemics span to larger portions of the nodes than in similar assortative ones.
This means that being anti-assortative is preferable for the spreading of knowledge and know-how in the productive system, making it more efficient.
It is worth noting that in a recent work \cite{EPJB12292} a social network similar to the one studied here, concerning the funding of FP7 (Seventh Framework Programme) European research projects, has been found to be anti-assortative as well, and conclusions close to the ones put foward here are drawn.
This could be an indication of some structural feature shared by graphs constructed starting from public funding programs, and we plan to further investigate this point in a future work.
Lastly, \emph{link efficiency} is a measure of traffic capacity within the network, representing how efficiently information can be transmitted along the graph.
This parameter takes the very high value $\xi=0.999$ in the PON R\&C network, which is a strong indicator of robustness for the relations between vertices, especially in a graph with small density $\rho=0.017$ as the one under examination.

The study of global properties of the PON R\&C graph is given as an example showing how network analysis provides a concrete way of examining the role of funded actors within a program, supporting its \emph{ex post} evaluation with the introduction of rather innovative indicators.
In particular, it is able to describe the structure of the productive system, highlighting the key nodes for network connectivity, or vertices that have a central role, through quantitative (thus evaluable) indicators, which for the present case are summarised in \hyperref[tab:ponrec_global_properties]{Tab.~\ref*{tab:ponrec_global_properties}}.
\begin{table}[h!]
	\begin{center}
		\footnotesize{
		\begin{tabular}{|c|c|}
			\hline\hline				&									\\
			\textbf{Property}			&	\textbf{Value}						\\
			\hline\hline				&									\\
			Radius					&	${\CMcal R}=3$					\\
			\hline					&									\\
			Diameter					&	${\CMcal D}=5$					\\
			\hline					&									\\
			Density					&	$\rho=0.017$						\\
			\hline					&									\\
			Global clustering coefficient	&	${\CMcal C}=0.215$					\\
			\hline					&									\\
			Vertex connectivity			&	$V_{c}=1$							\\
			\hline					&									\\
			Edge connectivity			&	$E_{c}=1$							\\
			\hline					&									\\
			Average shortest path length	&	$L=2.532$~~$({\CMcal O}(\log\log N))$	\\
			\hline					&									\\
			Link efficiency				&	$\xi=0.999$						\\
			\hline					&									\\
			Assortativity coefficient		&	$r=-0.173$						\\
			\hline\hline
		\end{tabular}
		}
		\caption{Global properties of the PON R\&C network.}\label{tab:ponrec_global_properties}
	\end{center}
\end{table}

\subsection{Community structure}\label{subsec:community_structure}

\begin{figure}
	\begin{center}
		\includegraphics[scale=0.35]{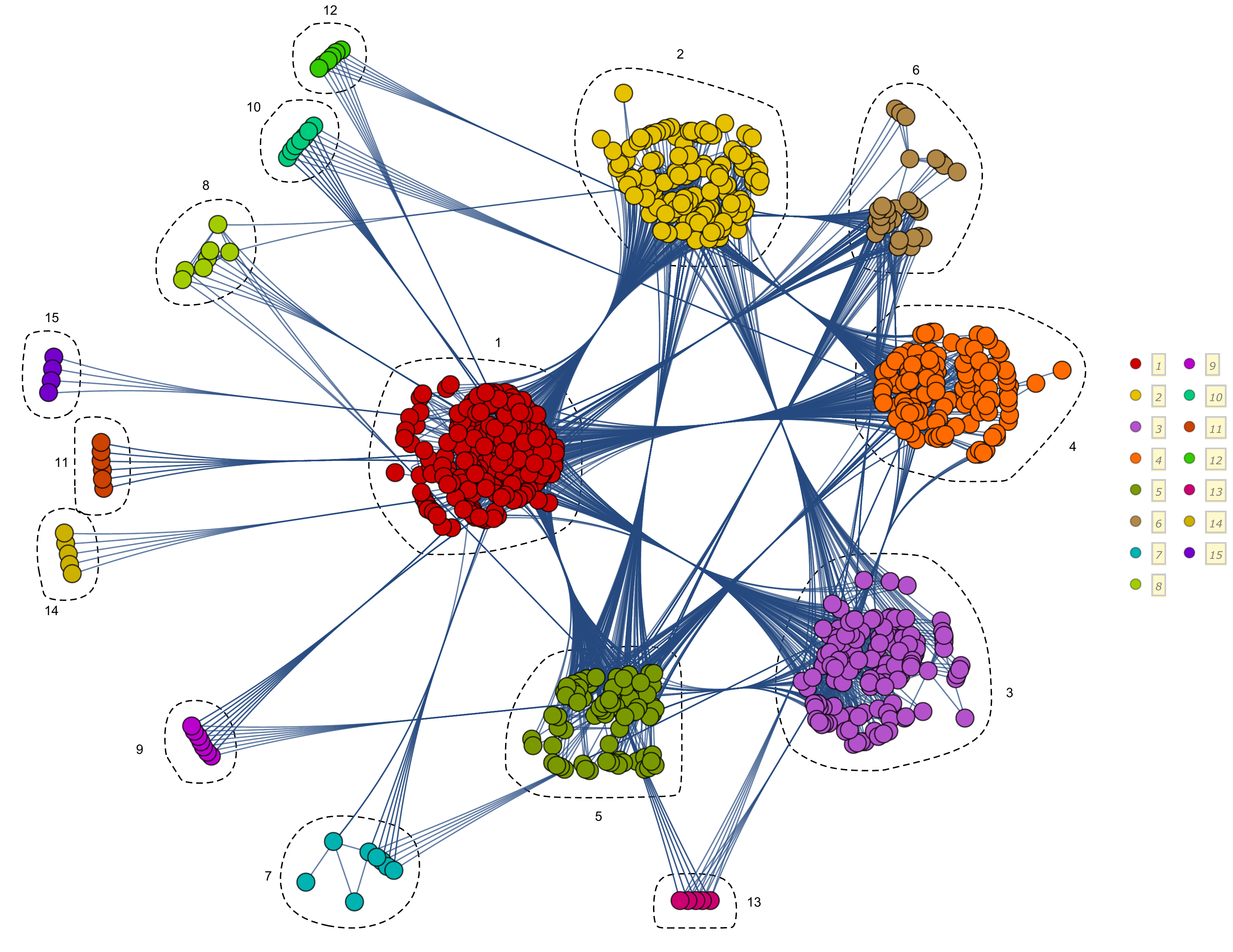}
		\caption{\small The community structure of the (giant component of the) PON R\&C network. 14 communities are highlighted, found with the Newman-Girvan algorithm.}
		\label{fig:pon_rec_communities}
	\end{center}
\end{figure}

The community structure of a graph is a global property, but a separate section is dedicated to it, since it has a special role, of particular importance for program evaluation.
In the PON R\&C network 15 communities are found with the Newman-Girvan algorithm \cite{PhysRevE.69.026113}, composed of 207, 136, 129, 113, 75, 29, 8, 7, 7, 7, 6, 6, 5, 5, and 4 vertices, respectively.
The algorithm consists in recursively removing from the graph the edges with the highest edge betweenness and recalculating the edge betweenness for the new graph obtained at each step.
This procedure generates a dendrogram of sets of communities, from which the set with largest modularity is chosen.
The community structure of the PON R\&C network is shown in \hyperref[fig:pon_rec_communities]{Fig.~\ref*{fig:pon_rec_communities}}.
The reason why it is so important is that it represents an unbiased way of discovering the existence of groups within a certain network of relationships, and highlighting such groups can be very important for the analysis of a productive system like the one described by the graph under examination.
The PON R\&C network shows strongly heterogeneous communities, with hugely populated groups and very small ones.
An important point, that can be interesting for an \emph{ex post} program evaluator, is that when communities grow in size, they tend to include important nodes.
For example, the biggest community, made of 186 vertices, include the CNR in it, which shows the record values for all the centralities, as stated in \hyperref[subsec:local_properties]{Sec.~\ref*{subsec:local_properties}}.

\begin{figure}
	\begin{center}
		\includegraphics[scale=0.4]{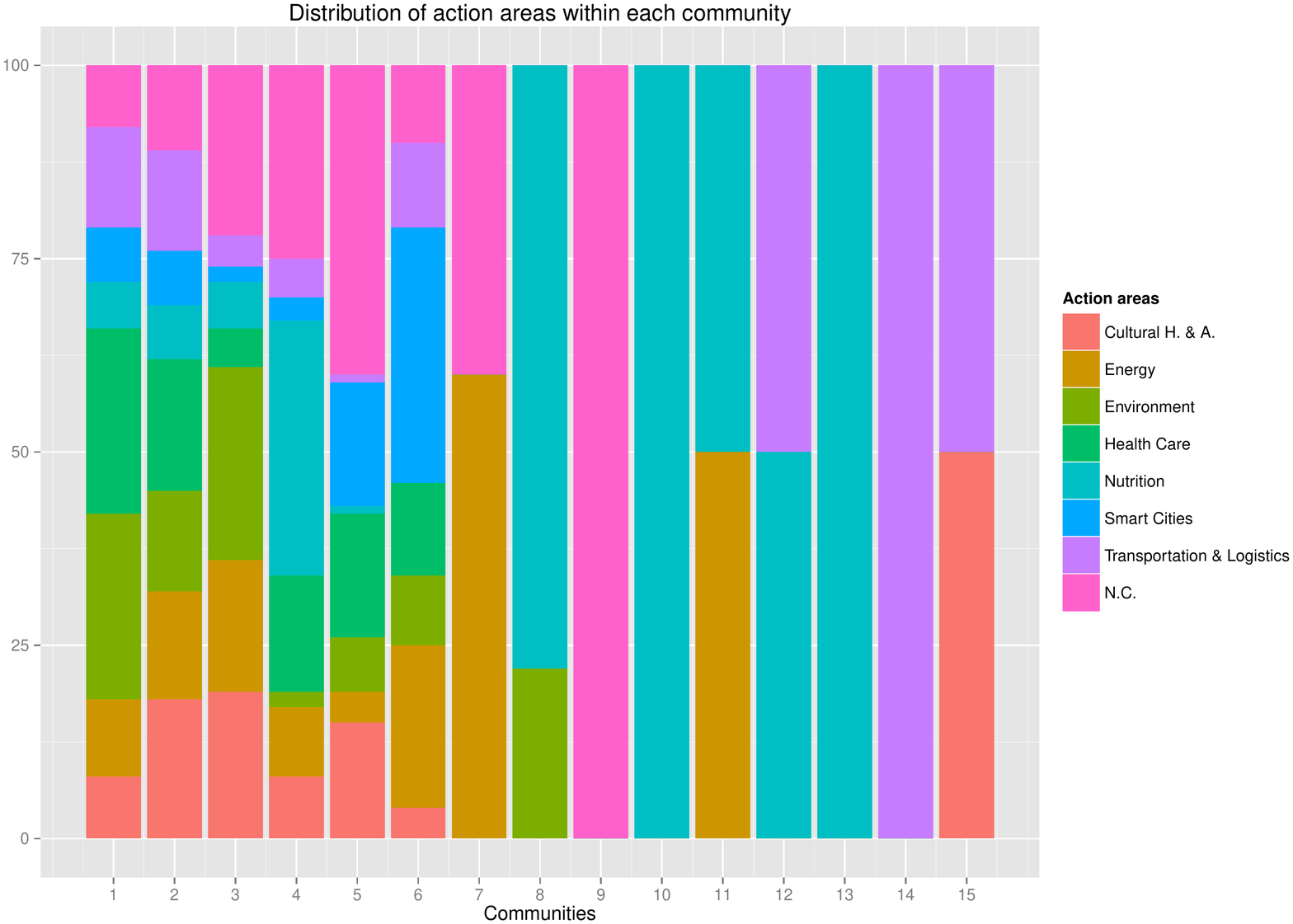}
		\caption{\small Distribution of action areas within each community.}
		\label{fig:action_areas_communities}
	\end{center}
\end{figure}

Moreover, comparing the community structure coming from network analysis with the one expected on the basis of external (economical, political, and social) considerations can enrich the evaluation by introducing a different point of view on the system under examination, not driven by ``human'' considerations, but purely mathematical in nature.
The distribution in percentage of action areas within each community is shown in \hyperref[fig:action_areas_communities]{Fig.~\ref*{fig:action_areas_communities}}.
It can be seen that the communities found with network analysis are not directly linked to action areas, at least the largest ones.
Other algorithms can be used to extract the community structure of a graph, like the leading eigenvector \cite{newman2006finding}, the multi-level modularity \cite{blondel2008fast} or the spin-glass \cite{reichardt2006statistical}, and refined methods can be used to which one gives the most significant results, like \emph{e.g.} a consensus analysis \cite{lancichinetti2012consensus}.
This kind of study is outside the purposes of the present paper, since it must be policy-driven, rather than research-driven, in order to be of interest for program evaluation.
We plan to develop similar analyses in future work.

\section{Conclusions and perspectives}\label{sec:conclusions}

In this paper we have used techniques borrowed from complex network analysis to evaluate the effects of a public funding program on the relations between the funded ``actors''.
The PON~R\&C program involves a large number of actors and is extended over a period of seven years (2007-2013).
The dataset is completely made of Open Data, and we have shown a way of concretely using information made available by Governments, in the spirit promoted by current global guidelines.
We have described the full process of knowledge management, from data acquisition, to cleaning, model building and querying.
The whole chain is data oriented and is focused in retaining every piece of available information, in order for the output of the analysis to show the highest possible accuracy.  

The processed PON~R\&C data have been used for complex network analysis, and the resulting network has 769 vertices and 4868 edges.
We have evaluated the most important centralities for each node, plus some relevant global properties of the graph.
The outcome of our analysis shows a dominant role of public (and, but less importantly, also private) research institutions within the Italian productive panorama, at least for the part portrayed by the program under examination.
Universities and research centres play the role of the ``glue'' for this particular program, \emph{i.e.} they are responsible of the connectedness of the network, and a failure involving some of them would be disruptive for the whole productive system.
This picture was partly expected, due to the way the program has been realised, since in many cases it was mandatory to involve public research institutions in the projects.
Nevertheless it can be useful to use such result as an \emph{ex-post} indicator.
Moreover, we have found that the PON~R\&C network is anti-assortative, an unusual feature of social networks, shared only with other cases involving FP7 public program of European research funding, preferable than the most common assortative mixing for the spreading of knowledge and know-how in the productive system and for its efficiency. 

We have shown that social network analysis can produce useful results for program evaluators, since it allows to consider, in a quantitative fashion, very important aspects that are usually ignored, due to common difficulties in quantifying them.
A mathematical description of the structure of relations generated by a national funding program is the example shown in this work.
Indicators such as vertex and edge centralities have been used to generate a ranking between the main actors involved in the program, as shown in \hyperref[fig:vertex_centralities]{Fig.~\ref*{fig:vertex_centralities}} and \hyperref[fig:edge_betweenness]{Fig.~\ref*{fig:edge_betweenness}}.
We hope that the procedure and the results described in the present paper can help opening interesting new perspectives from new indicators for decision and policy makers and program evaluators, providing them with an useful tool.  

Many possibilities are left open by the present work.
First of all, as mentioned in \hyperref[sec:dataset]{Sec.~\ref*{sec:dataset}}, around $\sim$78\% of the total budget is concentrated into $\sim$10\% of the funded projects.
This suggests that introducing information about the financial aspect into the network analysis could be interesting and meaningful for the evaluator.
This could be done in many different ways, from simple visualisation techniques in sociograms, like relating the size of the nodes to the total funding received by the actor, to more refined analysis, like defining weighted networks with weights related to budgets.
We plan to investigate these directions in future works.
Other planned future activities include the introduction of dynamical networks, involving the study of temporal series, refining of network analysis techniques, \emph{e.g.} by introducing different kinds of weighted networks and related features, and generalising the analysis extending it to different levels \cite{EPJB12292,PUAR:PUAR045}.
Moreover, the expected improvement in quality of Open Data (for example increasing the level of detail within public research institution, \emph{e.g.} discriminating among single departments rather than universities) could lead to many interesting improvements of the present analysis.

%%%%%%%%%%%%%%%%%%%%%%%%%%%%%%%%%%%%%%%%%%%%%%
%%
%% Backmatter begins here
%%
%%%%%%%%%%%%%%%%%%%%%%%%%%%%%%%%%%%%%%%%%%%%%%

\begin{backmatter}

\section*{Competing interests}
The authors declare that they have no competing interests.

\section*{Authors' contributions}
All authors designed research.
SN designed the network model.
ET designed the data model.
SN and ET prepared the data for network analysis. 
All authors analysed data. 
All authors read and approved the final manuscript.

\section*{Acknowledgements}
We thank N.~Coniglio for valuable suggestions and discussions.
SN thanks R.~Anglani and V.~Mariani for discussions.
SN, and NA acknowledge funding by the Italian MIUR grant \emph{PON PRISMA} Cod. PON04a2\_A.
ET acknowledges partial funding by the Italian MIUR grant \emph{PON PRISMA} Cod. PON04a2\_A.

%%%%%%%%%%%%%%%%%%%%%%%%%%%%%%%%%%%%%%%%%%%%%%%%%%%%%%%%%%%%%
%%                  The Bibliography
%%
%%  Bmc_mathpys.bst  will be used to
%%  create a .BBL file for submission. 
%%  After submission of the .TEX file,
%%  you will be prompted to submit your .BBL file.
%%
%%
%%  Note that the displayed Bibliography will not 
%%  necessarily be rendered by Latex exactly as specified
%%  in the online Instructions for Authors.
%%
%%%%%%%%%%%%%%%%%%%%%%%%%%%%%%%%%%%%%%%%%%%%%%%%%%%%%%%%%%%%%

% if your bibliography is in bibtex format, use those commands:

% or include bibliography directly:
%% BioMed_Central_Bib_Style_v1.01

\newcommand{\BMCxmlcomment}[1]{}

\BMCxmlcomment{

<refgrp>

<bibl id="B1">
  <title><p>{Mad Cows and Ecstasy: Chance and Choice in an Evidence-Based
  Society}</p></title>
  <aug>
    <au><snm>Smith</snm><fnm>AFM</fnm></au>
  </aug>
  <source>Journal of the Royal Statistical Society. Series A (Statistics in
  Society)</source>
  <publisher>Blackwell Publishing for the Royal Statistical Society</publisher>
  <pubdate>1996</pubdate>
  <volume>159</volume>
  <issue>3</issue>
  <fpage>367</fpage>
  <lpage>-383</lpage>
  <url>http://dx.doi.org/10.2307/2983324</url>
</bibl>

<bibl id="B2">
  <title><p>{What Works?: Evidence-based Policy and Practice in Public
  Services}</p></title>
  <aug>
    <au><snm>Nutley</snm><fnm>S.M.</fnm></au>
    <au><snm>Davies</snm><fnm>H.T.O.</fnm></au>
    <au><snm>Smith</snm><fnm>P.C.</fnm></au>
  </aug>
  <publisher>Policy Press</publisher>
  <pubdate>2000</pubdate>
</bibl>

<bibl id="B3">
  <title><p>{Evidence Based Policy: Whence it Came and where It's
  Going}</p></title>
  <aug>
    <au><snm>Solesbury</snm><fnm>W.</fnm></au>
    <au><snm>Policy</snm><fnm>EUCfEB</fnm></au>
    <au><cnm>Practice</cnm></au>
  </aug>
  <publisher>ESRC UK Centre for Evidence Based Policy and Practice</publisher>
  <pubdate>2001</pubdate>
</bibl>

<bibl id="B4">
  <title><p>{Social Science and the Evidence-based Policy Movement}</p></title>
  <aug>
    <au><snm>Young</snm><fnm>K</fnm></au>
    <au><snm>Ashby</snm><fnm>D</fnm></au>
    <au><snm>Boaz</snm><fnm>A</fnm></au>
    <au><snm>Grayson</snm><fnm>L</fnm></au>
  </aug>
  <source>Social Policy and Society</source>
  <pubdate>2002</pubdate>
  <volume>1</volume>
  <fpage>215</fpage>
  <lpage>-224</lpage>
  <url>http://journals.cambridge.org/article_S1474746402003068</url>
</bibl>

<bibl id="B5">
  <title><p>{Evidence-Based Policy-Making: The Elusive Search for Rational
  Public Administration}</p></title>
  <aug>
    <au><snm>Kay</snm><fnm>A</fnm></au>
  </aug>
  <source>Australian Journal of Public Administration</source>
  <publisher>Blackwell Publishing Asia</publisher>
  <pubdate>2011</pubdate>
  <volume>70</volume>
  <issue>3</issue>
  <fpage>236</fpage>
  <lpage>-245</lpage>
  <url>http://dx.doi.org/10.1111/j.1467-8500.2011.00728.x</url>
</bibl>

<bibl id="B6">
  <title><p>{Overseas and over here: policy transfer and evidence-based
  policy-making}</p></title>
  <aug>
    <au><snm>Legrand</snm><fnm>T</fnm></au>
  </aug>
  <source>Policy Studies</source>
  <pubdate>2012</pubdate>
  <volume>33</volume>
  <issue>4</issue>
  <fpage>329</fpage>
  <lpage>-348</lpage>
  <url>http://dx.doi.org/10.1080/01442872.2012.695945</url>
</bibl>

<bibl id="B7">
  <title><p>{Do Networks Really Work? A Framework for Evaluating Public-Sector
  Organizational Networks}</p></title>
  <aug>
    <au><snm>Provan</snm><fnm>KG</fnm></au>
    <au><snm>Milward</snm><fnm>HB</fnm></au>
  </aug>
  <source>Public Administration Review</source>
  <publisher>Blackwell Publishers Ltd.</publisher>
  <pubdate>2001</pubdate>
  <volume>61</volume>
  <issue>4</issue>
  <fpage>414</fpage>
  <lpage>-423</lpage>
  <url>http://dx.doi.org/10.1111/0033-3352.00045</url>
</bibl>

<bibl id="B8">
  <title><p>{Networks and Inter-Organizational ManagementChallenging, Steering,
  Evaluation, and the Role of Public Actors in Public Management}</p></title>
  <aug>
    <au><cnm>Ferlie</cnm></au>
    <au><cnm>Lynn</cnm></au>
    <au><cnm>Pollitt</cnm></au>
    <au><cnm>Klijn</cnm></au>
  </aug>
  <publisher>Oxford University Press</publisher>
  <url>http://www.oxfordhandbooks.com/10.1093/oxfordhb/9780199226443.001.0001/oxfordhb-9780199226443-e-12</url>
</bibl>

<bibl id="B9">
  <title><p>{Investigating the Potential of Using Social Network Analysis in
  Educational Evaluation}</p></title>
  <aug>
    <au><snm>Penuel</snm><fnm>WR</fnm></au>
    <au><snm>Sussex</snm><fnm>W</fnm></au>
    <au><snm>Korbak</snm><fnm>C</fnm></au>
    <au><snm>Hoadley</snm><fnm>C</fnm></au>
  </aug>
  <source>American Journal of Evaluation</source>
  <pubdate>2006</pubdate>
  <volume>27</volume>
  <issue>4</issue>
  <fpage>437</fpage>
  <lpage>-451</lpage>
  <url>http://aje.sagepub.com/content/27/4/437.abstract</url>
</bibl>

<bibl id="B10">
  <title><p>{Network Evaluation from the Everyday Life Perspective: A Tool for
  Capacity-Building and Voice}</p></title>
  <aug>
    <au><snm>Horelli</snm><fnm>L</fnm></au>
  </aug>
  <source>Evaluation</source>
  <pubdate>2009</pubdate>
  <volume>15</volume>
  <issue>2</issue>
  <fpage>205</fpage>
  <lpage>-223</lpage>
  <url>http://evi.sagepub.com/content/15/2/205.abstract</url>
</bibl>

<bibl id="B11">
  <title><p>{Networks In Evaluation}</p></title>
  <aug>
    <au><snm>Ploszaj</snm><fnm>A.</fnm></au>
  </aug>
  <source>{Evaluating the effects of regional interventions. A look beyond
  current Structural Funds practice}</source>
  <publisher>Warszawa: MRR</publisher>
  <editor>Olejniczak, K. and Kozak, M. and Bienias, S.</editor>
  <pubdate>2011</pubdate>
</bibl>

<bibl id="B12">
  <title><p>{Network of participants in European research: accepted versus
  rejected proposals}</p></title>
  <aug>
    <au><snm>Tsouchnika</snm><fnm>M</fnm></au>
    <au><snm>Argyrakis</snm><fnm>P</fnm></au>
  </aug>
  <source>Eur. Phys. J. B</source>
  <pubdate>2014</pubdate>
  <volume>87</volume>
  <issue>12</issue>
  <fpage>292</fpage>
  <url>http://dx.doi.org/10.1140/epjb/e2014-50450-4</url>
</bibl>

<bibl id="B13">
  <title><p>{Agenda Digitale Puglia 2020}</p></title>
  <aug>
    <au><cnm>{Regione Puglia, Area Politiche per il Lavoro Sviluppo e
  Innovazione Servizio ricerca Industriale e Innovazione}</cnm></au>
    <au><cnm>{InnovaPuglia S.p.A.}</cnm></au>
  </aug>
  <source>Bollettino Ufficiale della Regione Puglia (BURP)</source>
  <pubdate>2014</pubdate>
  <issue>128</issue>
  <fpage>33423</fpage>
  <lpage>-33502</lpage>
</bibl>

<bibl id="B14">
  <title><p>{Linked data-the story so far}</p></title>
  <aug>
    <au><snm>Bizer</snm><fnm>C</fnm></au>
    <au><snm>Heath</snm><fnm>T</fnm></au>
    <au><snm>Berners Lee</snm><fnm>T</fnm></au>
  </aug>
  <source>International journal on semantic web and information
  systems</source>
  <pubdate>2009</pubdate>
  <volume>5</volume>
  <issue>3</issue>
  <fpage>1</fpage>
  <lpage>-22</lpage>
</bibl>

<bibl id="B15">
  <title><p>{Fundamentals of data warehouses}</p></title>
  <aug>
    <au><snm>Lenzerini</snm><fnm>M</fnm></au>
    <au><snm>Vassiliou</snm><fnm>Y</fnm></au>
    <au><snm>Vassiliadis</snm><fnm>P</fnm></au>
    <au><snm>Jarke</snm><fnm>M</fnm></au>
  </aug>
  <publisher>Springer</publisher>
  <pubdate>2003</pubdate>
</bibl>

<bibl id="B16">
  <title><p>{Foundations of databases}</p></title>
  <aug>
    <au><snm>Abiteboul</snm><fnm>S</fnm></au>
    <au><snm>Hull</snm><fnm>R</fnm></au>
    <au><snm>Vianu</snm><fnm>V</fnm></au>
  </aug>
  <publisher>Addison-Wesley Reading</publisher>
  <pubdate>1995</pubdate>
  <volume>8</volume>
</bibl>

<bibl id="B17">
  <title><p>{Five stars of Linked Data vocabulary use}</p></title>
  <aug>
    <au><snm>Janowicz</snm><fnm>K</fnm></au>
    <au><snm>Hitzler</snm><fnm>P</fnm></au>
    <au><snm>Adams</snm><fnm>B</fnm></au>
    <au><snm>Kolas</snm><fnm>D</fnm></au>
    <au><snm>{Vardeman II}</snm><fnm>C</fnm></au>
  </aug>
  <source>Semantic Web</source>
  <publisher>IOS Press</publisher>
  <pubdate>2014</pubdate>
  <volume>5</volume>
  <issue>3</issue>
  <fpage>173</fpage>
  <lpage>-176</lpage>
</bibl>

<bibl id="B18">
  <title><p>{Scientific collaboration networks. I. Network construction and
  fundamental results}</p></title>
  <aug>
    <au><snm>Newman</snm><fnm>M. E. J.</fnm></au>
  </aug>
  <source>Phys. Rev. E</source>
  <publisher>American Physical Society</publisher>
  <pubdate>2001</pubdate>
  <volume>64</volume>
  <fpage>016131</fpage>
  <url>http://link.aps.org/doi/10.1103/PhysRevE.64.016131</url>
</bibl>

<bibl id="B19">
  <title><p>{Scientific collaboration networks. II. Shortest paths, weighted
  networks, and centrality}</p></title>
  <aug>
    <au><snm>Newman</snm><fnm>M. E. J.</fnm></au>
  </aug>
  <source>Phys. Rev. E</source>
  <publisher>American Physical Society</publisher>
  <pubdate>2001</pubdate>
  <volume>64</volume>
  <fpage>016132</fpage>
  <url>http://link.aps.org/doi/10.1103/PhysRevE.64.016132</url>
</bibl>

<bibl id="B20">
  <title><p>{Evaluating the effects of regional interventions. A look beyond
  current Structural Funds practice}</p></title>
  <aug>
    <au><snm>Olejniczak</snm><fnm>K</fnm></au>
    <au><snm>Bienias</snm><fnm>S</fnm></au>
    <au><snm>Kozak</snm><fnm>M</fnm></au>
  </aug>
  <publisher>Ministry of Regional Development, Republic of Poland</publisher>
  <pubdate>2012</pubdate>
</bibl>

<bibl id="B21">
  <title><p>{Complex networks: Structure and dynamics}</p></title>
  <aug>
    <au><snm>Boccaletti</snm><fnm>S.</fnm></au>
    <au><snm>Latora</snm><fnm>V.</fnm></au>
    <au><snm>Moreno</snm><fnm>Y.</fnm></au>
    <au><snm>Chavez</snm><fnm>M.</fnm></au>
    <au><snm>Hwang</snm><fnm>D. U.</fnm></au>
  </aug>
  <source>Physics Reports</source>
  <pubdate>2006</pubdate>
  <volume>424</volume>
  <issue>4-5</issue>
  <fpage>175</fpage>
  <lpage>-308</lpage>
  <url>http://dx.doi.org/10.1016/j.physrep.2005.10.009</url>
</bibl>

<bibl id="B22">
  <title><p>{Statistical mechanics of complex networks}</p></title>
  <aug>
    <au><snm>Albert</snm><fnm>R</fnm></au>
    <au><snm>Barab{\'a}si</snm><fnm>AL</fnm></au>
  </aug>
  <source>Rev. Mod. Phys.</source>
  <publisher>American Physical Society</publisher>
  <pubdate>2002</pubdate>
  <volume>74</volume>
  <fpage>47</fpage>
  <lpage>-97</lpage>
  <url>http://link.aps.org/doi/10.1103/RevModPhys.74.47</url>
</bibl>

<bibl id="B23">
  <title><p>{The anatomy of a large-scale hypertextual Web search
  engine}</p></title>
  <aug>
    <au><snm>Brin</snm><fnm>S</fnm></au>
    <au><snm>Page</snm><fnm>L</fnm></au>
  </aug>
  <source>Computer Networks and \{ISDN\} Systems</source>
  <pubdate>1998</pubdate>
  <volume>30</volume>
  <issue>1--7</issue>
  <fpage>107</fpage>
  <lpage>-117</lpage>
  <url>http://www.sciencedirect.com/science/article/pii/S016975529800110X</url>
  <note>Proceedings of the Seventh International World Wide Web
  Conference</note>
</bibl>

<bibl id="B24">
  <title><p>{A Set of Measures of Centrality Based on Betweenness}</p></title>
  <aug>
    <au><snm>Freeman</snm><fnm>LC</fnm></au>
  </aug>
  <source>Sociometry</source>
  <publisher>American Sociological Association</publisher>
  <pubdate>1977</pubdate>
  <volume>40</volume>
  <issue>1</issue>
  <fpage>pp.35</fpage>
  <lpage>-41</lpage>
  <url>http://www.jstor.org/stable/3033543</url>
</bibl>

<bibl id="B25">
  <title><p>{Method for node ranking in a linked database}</p></title>
  <aug>
    <au><snm>Page</snm><fnm>L.</fnm></au>
  </aug>
  <publisher>Google Patents</publisher>
  <pubdate>2001</pubdate>
  <url>http://www.google.com/patents/US6285999</url>
  <note>US Patent 6,285,999</note>
</bibl>

<bibl id="B26">
  <title><p>{Annotating links in a document based on the ranks of documents
  pointed to by the links}</p></title>
  <aug>
    <au><snm>Page</snm><fnm>L.</fnm></au>
  </aug>
  <publisher>Google Patents</publisher>
  <pubdate>2011</pubdate>
  <url>http://www.google.com/patents/US7908277</url>
  <note>US Patent 7,908,277</note>
</bibl>

<bibl id="B27">
  <title><p>{Scoring documents in a linked database}</p></title>
  <aug>
    <au><snm>Page</snm><fnm>L.</fnm></au>
  </aug>
  <publisher>Google Patents</publisher>
  <pubdate>2014</pubdate>
  <url>http://www.google.com/patents/US8725726</url>
  <note>US Patent 8,725,726</note>
</bibl>

<bibl id="B28">
  <title><p>{Emergence of Scaling in Random Networks}</p></title>
  <aug>
    <au><snm>Barab{\'a}si</snm><fnm>AL</fnm></au>
    <au><snm>Albert</snm><fnm>R</fnm></au>
  </aug>
  <source>Science</source>
  <pubdate>1999</pubdate>
  <volume>286</volume>
  <issue>5439</issue>
  <fpage>509</fpage>
  <lpage>-512</lpage>
  <url>http://www.sciencemag.org/content/286/5439/509.abstract</url>
</bibl>

<bibl id="B29">
  <title><p>{Topology of Evolving Networks: Local Events and
  Universality}</p></title>
  <aug>
    <au><snm>Albert</snm><fnm>R</fnm></au>
    <au><snm>Barab{\'a}si</snm><fnm>AL</fnm></au>
  </aug>
  <source>Phys. Rev. Lett.</source>
  <publisher>American Physical Society</publisher>
  <pubdate>2000</pubdate>
  <volume>85</volume>
  <fpage>5234</fpage>
  <lpage>-5237</lpage>
  <url>http://link.aps.org/doi/10.1103/PhysRevLett.85.5234</url>
</bibl>

<bibl id="B30">
  <title><p>{First- and second-order phase transitions in scale-free
  networks}</p></title>
  <aug>
    <au><snm>Igl{\'o}i</snm><fnm>F</fnm></au>
    <au><snm>Turban</snm><fnm>L</fnm></au>
  </aug>
  <source>Phys. Rev. E</source>
  <publisher>American Physical Society</publisher>
  <pubdate>2002</pubdate>
  <volume>66</volume>
  <fpage>036140</fpage>
  <url>http://link.aps.org/doi/10.1103/PhysRevE.66.036140</url>
</bibl>

<bibl id="B31">
  <title><p>{Nonlinear Programming: Analysis and Methods}</p></title>
  <aug>
    <au><snm>Avriel</snm><fnm>M.</fnm></au>
  </aug>
  <publisher>Dover Publications</publisher>
  <series><title><p>{Dover Books on Computer Science
  Series}</p></title></series>
  <pubdate>2003</pubdate>
</bibl>

<bibl id="B32">
  <title><p>{Statistical methods in experimental physics}</p></title>
  <aug>
    <au><snm>Eadie</snm><fnm>WT</fnm></au>
    <au><snm>Drijard</snm><fnm>D</fnm></au>
    <au><snm>James</snm><fnm>FE</fnm></au>
    <au><snm>Roos</snm><fnm>M</fnm></au>
    <au><snm>Sadoulet</snm><fnm>B</fnm></au>
  </aug>
  <publisher>Amsterdam: North-Holland</publisher>
  <pubdate>1971</pubdate>
</bibl>

<bibl id="B33">
  <title><p>{Collective dynamics of /`small-world/' networks}</p></title>
  <aug>
    <au><snm>Watts</snm><fnm>DJ</fnm></au>
    <au><snm>Strogatz</snm><fnm>SH</fnm></au>
  </aug>
  <source>Nature</source>
  <pubdate>1998</pubdate>
  <volume>393</volume>
  <issue>6684</issue>
  <fpage>440</fpage>
  <lpage>-442</lpage>
  <url>http://dx.doi.org/10.1038/30918</url>
</bibl>

<bibl id="B34">
  <title><p>{The small world problem}</p></title>
  <aug>
    <au><snm>Milgram</snm><fnm>S.</fnm></au>
  </aug>
  <source>Psychology Today</source>
  <pubdate>1967</pubdate>
  <volume>2</volume>
  <issue>1</issue>
  <fpage>60</fpage>
  <lpage>-67</lpage>
</bibl>

<bibl id="B35">
  <title><p>{The average distances in random graphs with given expected
  degrees}</p></title>
  <aug>
    <au><snm>Chung</snm><fnm>F</fnm></au>
    <au><snm>Lu</snm><fnm>L</fnm></au>
  </aug>
  <source>Proceedings of the National Academy of Sciences</source>
  <pubdate>2002</pubdate>
  <volume>99</volume>
  <issue>25</issue>
  <fpage>15879</fpage>
  <lpage>-15882</lpage>
  <url>http://www.pnas.org/content/99/25/15879.abstract</url>
</bibl>

<bibl id="B36">
  <title><p>{Mean-field theory for scale-free random networks}</p></title>
  <aug>
    <au><snm>Barab{\'a}si</snm><fnm>AL</fnm></au>
    <au><snm>Albert</snm><fnm>R</fnm></au>
    <au><snm>Jeong</snm><fnm>H</fnm></au>
  </aug>
  <source>Physica A: Statistical Mechanics and its Applications</source>
  <pubdate>1999</pubdate>
  <volume>272</volume>
  <issue>1--2</issue>
  <fpage>173</fpage>
  <lpage>-187</lpage>
  <url>http://www.sciencedirect.com/science/article/pii/S0378437199002915</url>
</bibl>

<bibl id="B37">
  <title><p>{Evolution of networks}</p></title>
  <aug>
    <au><snm>Dorogovtsev</snm><fnm>S. N.</fnm></au>
    <au><snm>Mendes</snm><fnm>J. F. F.</fnm></au>
  </aug>
  <source>Advances in Physics</source>
  <pubdate>2002</pubdate>
  <volume>51</volume>
  <issue>4</issue>
  <fpage>1079</fpage>
  <lpage>-1187</lpage>
  <url>http://dx.doi.org/10.1080/00018730110112519</url>
</bibl>

<bibl id="B38">
  <title><p>{Connectivity of Growing Random Networks}</p></title>
  <aug>
    <au><snm>Krapivsky</snm><fnm>P. L.</fnm></au>
    <au><snm>Redner</snm><fnm>S.</fnm></au>
    <au><snm>Leyvraz</snm><fnm>F.</fnm></au>
  </aug>
  <source>Phys. Rev. Lett.</source>
  <publisher>American Physical Society</publisher>
  <pubdate>2000</pubdate>
  <volume>85</volume>
  <fpage>4629</fpage>
  <lpage>-4632</lpage>
  <url>http://link.aps.org/doi/10.1103/PhysRevLett.85.4629</url>
</bibl>

<bibl id="B39">
  <title><p>{Assortative Mixing in Networks}</p></title>
  <aug>
    <au><snm>Newman</snm><fnm>M. E. J.</fnm></au>
  </aug>
  <source>Phys. Rev. Lett.</source>
  <publisher>American Physical Society</publisher>
  <pubdate>2002</pubdate>
  <volume>89</volume>
  <fpage>208701</fpage>
  <url>http://link.aps.org/doi/10.1103/PhysRevLett.89.208701</url>
</bibl>

<bibl id="B40">
  <title><p>{Mixing patterns in networks}</p></title>
  <aug>
    <au><snm>Newman</snm><fnm>M. E. J.</fnm></au>
  </aug>
  <source>Phys. Rev. E</source>
  <publisher>American Physical Society</publisher>
  <pubdate>2003</pubdate>
  <volume>67</volume>
  <fpage>026126</fpage>
  <url>http://link.aps.org/doi/10.1103/PhysRevE.67.026126</url>
</bibl>

<bibl id="B41">
  <title><p>{Finding and evaluating community structure in
  networks}</p></title>
  <aug>
    <au><snm>Newman</snm><fnm>M. E. J.</fnm></au>
    <au><snm>Girvan</snm><fnm>M.</fnm></au>
  </aug>
  <source>Phys. Rev. E</source>
  <publisher>American Physical Society</publisher>
  <pubdate>2004</pubdate>
  <volume>69</volume>
  <fpage>026113</fpage>
  <url>http://link.aps.org/doi/10.1103/PhysRevE.69.026113</url>
</bibl>

<bibl id="B42">
  <title><p>{Finding community structure in networks using the eigenvectors of
  matrices}</p></title>
  <aug>
    <au><snm>Newman</snm><fnm>M. E. J.</fnm></au>
  </aug>
  <source>Phys. Rev. E</source>
  <publisher>American Physical Society</publisher>
  <pubdate>2006</pubdate>
  <volume>74</volume>
  <fpage>036104</fpage>
  <url>http://link.aps.org/doi/10.1103/PhysRevE.74.036104</url>
</bibl>

<bibl id="B43">
  <title><p>{Fast unfolding of communities in large networks}</p></title>
  <aug>
    <au><snm>Blondel</snm><fnm>VD</fnm></au>
    <au><snm>Guillaume</snm><fnm>JL</fnm></au>
    <au><snm>Lambiotte</snm><fnm>R</fnm></au>
    <au><snm>Lefebvre</snm><fnm>E</fnm></au>
  </aug>
  <source>Journal of Statistical Mechanics: Theory and Experiment</source>
  <pubdate>2008</pubdate>
  <volume>2008</volume>
  <issue>10</issue>
  <fpage>P10008</fpage>
  <url>http://stacks.iop.org/1742-5468/2008/i=10/a=P10008</url>
</bibl>

<bibl id="B44">
  <title><p>{Statistical mechanics of community detection}</p></title>
  <aug>
    <au><snm>Reichardt</snm><fnm>J</fnm></au>
    <au><snm>Bornholdt</snm><fnm>S</fnm></au>
  </aug>
  <source>Phys. Rev. E</source>
  <publisher>American Physical Society</publisher>
  <pubdate>2006</pubdate>
  <volume>74</volume>
  <fpage>016110</fpage>
  <url>http://link.aps.org/doi/10.1103/PhysRevE.74.016110</url>
</bibl>

<bibl id="B45">
  <title><p>{Consensus clustering in complex networks}</p></title>
  <aug>
    <au><cnm>{Lancichinetti Andrea}</cnm></au>
    <au><cnm>{Fortunato Santo}</cnm></au>
  </aug>
  <source>Sci. Rep.</source>
  <publisher>Macmillan Publishers Limited. All rights reserved</publisher>
  <pubdate>2012</pubdate>
  <volume>2</volume>
  <url>http://www.nature.com/srep/2012/120327/srep00336/abs/srep00336.html\#supplementary-information</url>
  <note>10.1038/srep00336</note>
</bibl>

</refgrp>
} % end of \BMCxmlcomment
%%%%%%%%%%%%%%%%%%%%%%%%%%%%%%%%%%%
%%
%% Figures
%%
%% NB: this is for captions and
%% Titles. All graphics must be
%% submitted separately and NOT
%% included in the Tex document
%%
%%%%%%%%%%%%%%%%%%%%%%%%%%%%%%%%%%%

%%
%% Do not use \listoffigures as most will included as separate files

%\section*{Figures}
%  \begin{figure}[h!]
%  \caption{\csentence{Sample figure title.}
%      A short description of the figure content
%      should go here.}
%      \end{figure}
%
%\begin{figure}[h!]
%  \caption{\csentence{Sample figure title.}
%      Figure legend text.}
%      \end{figure}
%
%%%%%%%%%%%%%%%%%%%%%%%%%%%%%%%%%%%%
%%%
%%% Tables
%%%
%%%%%%%%%%%%%%%%%%%%%%%%%%%%%%%%%%%%
%
%%% Use of \listoftables is discouraged.
%%%
%\section*{Tables}
%\begin{table}[h!]
%\caption{Sample table title. This is where the description of the table should go.}
%      \begin{tabular}{cccc}
%        \hline
%           & B1  &B2   & B3\\ \hline
%        A1 & 0.1 & 0.2 & 0.3\\
%        A2 & ... & ..  & .\\
%        A3 & ..  & .   & .\\ \hline
%      \end{tabular}
%\end{table}

%%%%%%%%%%%%%%%%%%%%%%%%%%%%%%%%%%%
%%
%% Additional Files
%%
%%%%%%%%%%%%%%%%%%%%%%%%%%%%%%%%%%%

%\section*{Additional Files}
%  \subsection*{Additional file 1 --- Sample additional file title}
%    Additional file descriptions text (including details of how to
%    view the file, if it is in a non-standard format or the file extension).  This might
%    refer to a multi-page table or a figure.
%
%  \subsection*{Additional file 2 --- Sample additional file title}
%    Additional file descriptions text.

\end{backmatter}
\end{document}